\documentclass{aa}  %% For A&A format
\usepackage{graphicx}
\usepackage{txfonts}
\newcommand{\GILDAS}{\texttt{GILDAS}}

\newcommand{\IRAM}{\textrm{IRAM}}

\newcommand{\PdBI}{\textrm{PdBI}}
\newcommand{\CSO}{\textrm{\CSO}}

\newcommand{\ie} {{\em i.e.}}
\newcommand{\eg} {{\em e.g.}}

\newcommand{\thCO} {\mbox{$^{13}$CO}}       % 13CO
\newcommand{\twCO} {\mbox{$^{12}$CO}}       % 12CO
\newcommand{\CeiO} {\mbox{C$^{18}$O}}       % C18O
\newcommand{\twCS} {\mbox{$^{12}$CS}}       % 12CS
\newcommand{\tfSO} {\mbox{$^{34}$SO}}   % 34SO

\newcommand{\lCCCH}{\mbox{l-C$_3$H}}
\newcommand{\Jone}{\mbox{$J$=1--0}}
\newcommand{\Jtwo}{\mbox{$J$=2--1}}

\newcommand{\emm}[1]{\ensuremath{#1}}   % Ensures math mode.
\newcommand{\emr}[1]{\emm{\mathrm{#1}}} % Uses math roman fonts.
\newcommand{\unit}[1]{\emm{\, \emr{#1}}}

\newcommand{\mm}  {\unit{mm}}
\newcommand{\m }  {\unit{m}}

\newcommand{\Kkms}{\unit{K\,km\,s^{-1}}}
\newcommand{\MHz} {\unit{MHz}}

\newcommand{\pc}    {\unit{pc}}

\newcommand{\kms}   {\unit{km\,s^{-1}}}

\renewcommand{\deg}{\emm{^\circ}}

\newcommand{\TabObs}{%
  \begin{table*}
    \caption{Observation parameters.}
    \begin{center}
      \begin{tabular}{lrrc}
        \hline
        & \multicolumn{2}{c}{Phase center} & Number of fields \\
        Mosaic 1 &  $\alpha_{2000} = 05^h40^m54.27^s $ & $ \delta_{2000} =
        -02^\circ 28' 00''$ & 7 \\
        Mosaic 2 &  $\alpha_{2000} = 05^h40^m53.00^s $ & $ \delta_{2000} =
        -02^\circ 28' 00'' $ & 4 \\
        \hline
      \end{tabular}
      \medskip{}
      \begin{tabular}{lrcccr}
        \hline
        Molecule \& Line & Frequency  &   Beam & PA     & Noise$^{a}$ & \multicolumn{1}{c}{Obs. date} \\
        & (GHz)        & (arcsec) & ($\deg$) & (\Kkms{})     &     \\
        \hline
        Mosaic 1 \\
        \twCS{} \Jtwo{} &  97.981 & $3.65 \times 3.34$ & 48 & 1.2$\times 10^{-1}$ & Aug. \& Oct. 2004 and Mar. 2005 \\
        \CeiO{} \Jtwo{} & 219.560 & $6.54 \times 4.31$ & 65 & 9.8$\times 10^{-2}$ & Mar. 2003 \\
        \hline
        Mosaic 2 \\
        \twCO{} \Jone{} & 115.271 & $5.95 \times 5.00$ & 65 & 1.2$\times 10^{-1}$ & Nov. 1999 \\
        \twCO{} \Jtwo{} & 230.538 & $2.97 \times 2.47$ & 66 & 1.7$\times 10^{-1}$ & Nov. 1999 \\
        \hline
      \end{tabular}
    \end{center}
    $^{a}$ The noise values quoted here are the noises at the mosaic center
    (Mosaic noise is inhomogeneous due to primary beam correction; it 
    steeply increases at the mosaic edges). Those noise values have been 
    computed in 1\kms{} velocity bin.
    \label{tab:pdb}
  \end{table*}}

\newcommand{\TabFlux}{%
  \begin{table}
    \centering
    \caption{Calibrator fluxes in Jy.}
    \begin{tabular}{lrrrr}
      \hline
      Date & \multicolumn{2}{c}{B0420$-$014} & \multicolumn{2}{c}{B0607$-$157}\\
                 & 3\mm & 1\mm & 3\mm & 1\mm \\
      \hline
      20.08.2004 &  3.4 &  2.9 &  1.4 & 0.93 \\
      04.10.2004 &  3.4 &  2.9 &  1.6 & 0.90 \\
      27.02.2005 &  3.5 &  2.9 &  1.6 & 0.90 \\
      02.03.2005 &  3.2 &  2.3 &  1.6 & 0.89 \\
      12.03.2005 &  3.2 &  2.3 &  1.6 & 0.90 \\
      13.03.2005 &  3.2 &  2.3 &  1.6 & 1.00 \\
      \hline
    \end{tabular}
    \label{tab:fluxes}
  \end{table}}

\newcommand{\FigMaps}{%
  \begin{figure*}
    \centering \includegraphics[width=0.85\hsize{}]{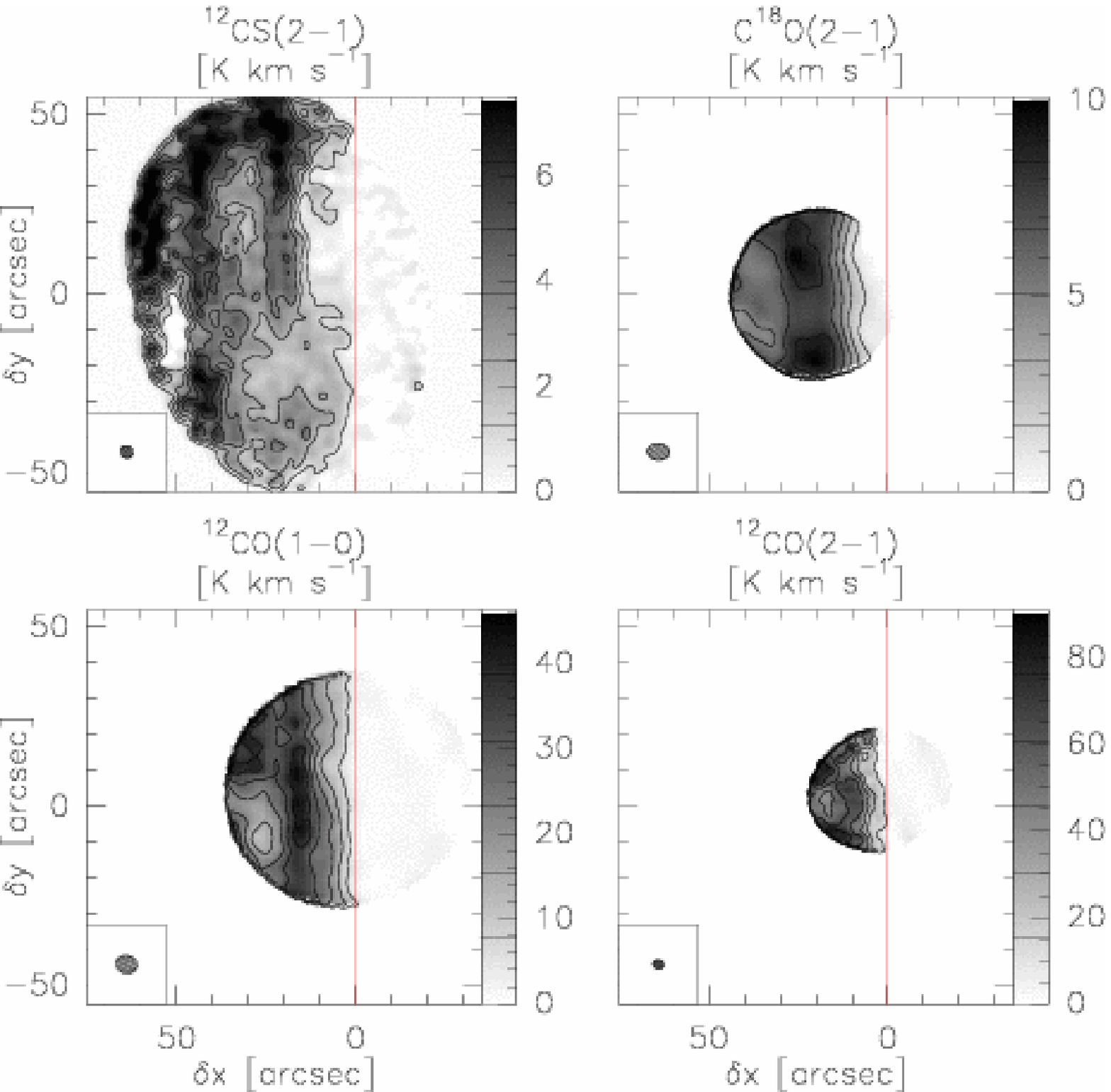}
    \caption{Integrated emission maps obtained with the Plateau de Bure 
      Interferometer. The center of all maps has been set to the mosaic
      phase center: RA(2000) = 05h40m54.27s, Dec(2000) =
      -02\deg{}28'00$''$.  The map size is $110'' \times 110''$, with ticks
      drawn every $10''$.  The synthesized beam is plotted in the bottom
      left corner.  The emission of all lines is integrated between
      10.1 and 11.1~\kms{}.  Values of contour level are shown on each
      image wedge. 
      The four panels are shown in a coordinate
      system adapted to the source: \ie{} maps have been rotated by
      14\deg{} counter--clockwise around the image center to bring the
      exciting star direction in the horizontal direction as this eases the
      comparison of the PDR models. Maps have also been
      horizontally shifted by $20''$ to set the horizontal zero at the PDR
      edge,  delineated by the vertical line.}
    \label{fig:maps1}
  \end{figure*}}

\newcommand{\FigMeanCuts}{%
  \begin{figure}[hb]
    \centering %
    \includegraphics[width=7cm]{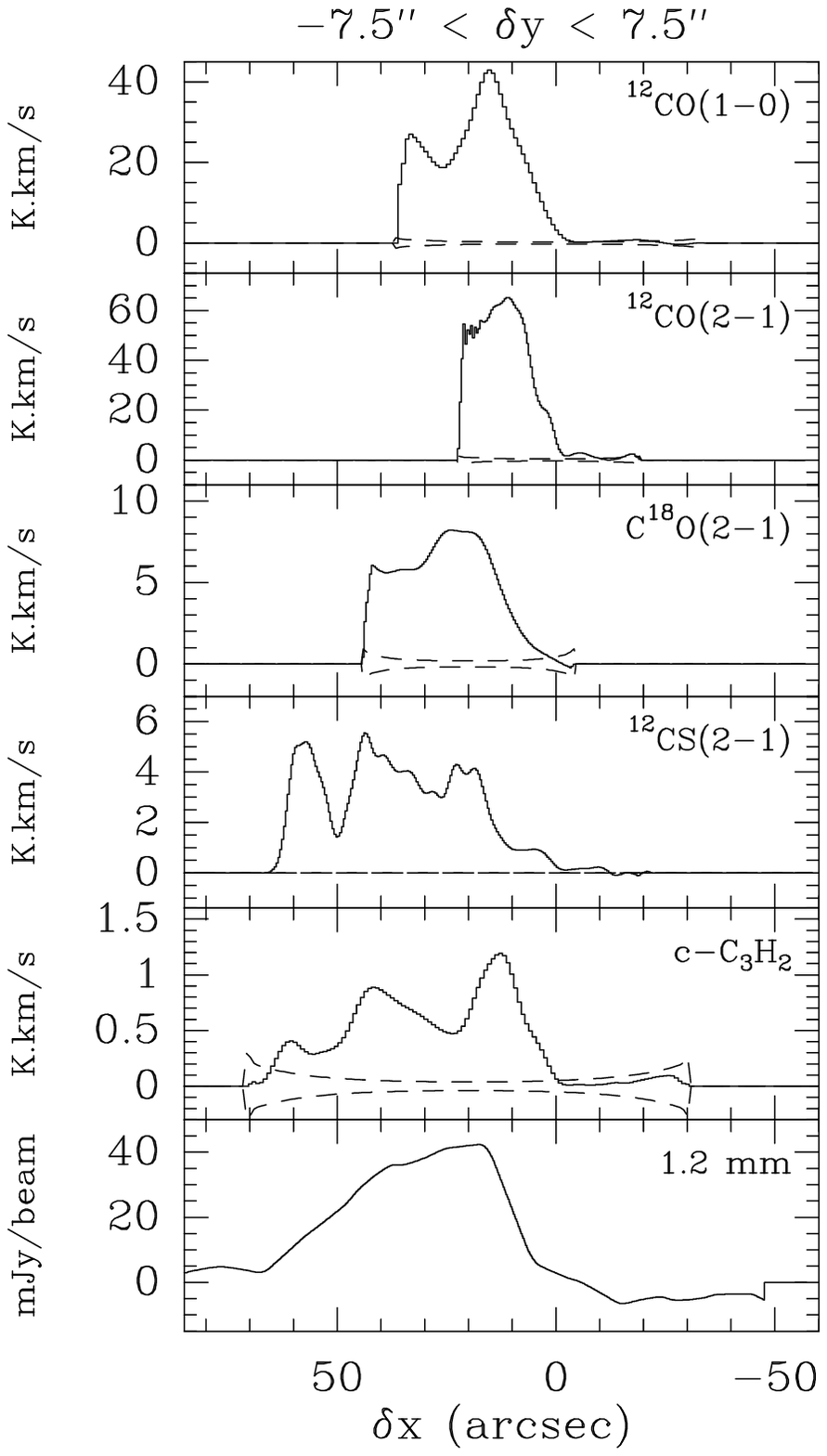}
    \caption{Emission profiles along the exciting star direction (PA =
      -104\deg{} in the equatorial coordinate system). To improve the
      signal--to--noise ratio,  emission profiles have been integrated
      along the perpendicular direction between $-7.5'' < \delta y <
      +7.5''$. We show from top to bottom \twCO{} \Jtwo{}, \twCO{} \Jone{},
      \CeiO{} \Jtwo{}, \twCS{} \Jtwo{}, $c$--C$_3$H$_2$ 2$_{12}$--1$_{01}$
      and 1.2~mm dust continuum emission (Pety et al. 2005a). The 
      3$\sigma$ noise level is
      indicated by the dashed lines. It rises at the cut edges due to the
      primary beam correction.  Note that the fields of view of the \twCO{}
      and \CeiO{} data are smaller than the field of view of the \twCS{}
      data because of the smaller mosaic size and/or the higher frequency.}
    \label{fig:cuts:mean}
  \end{figure}}

\newcommand{\FigSpectraCuts}{%
  \begin{figure*}
    \centering %
    \includegraphics[width=0.95\hsize{}]{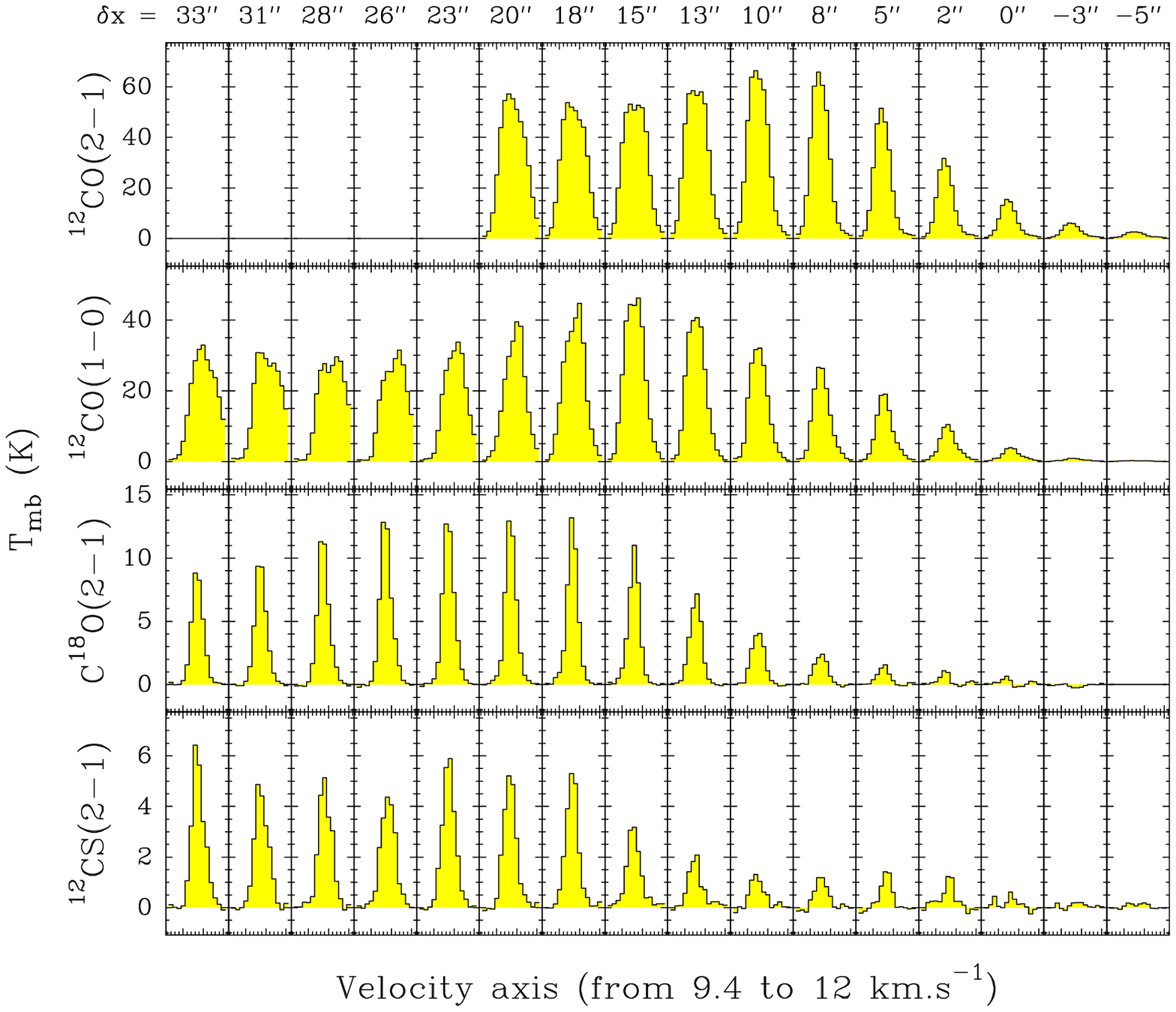}
    \caption{Spectra along the direction of the exciting star at 
      $\delta y = 0''$. \twCO{} \Jone{}, \CeiO{} \Jtwo{} and \twCS{}
      \Jtwo{} spectra cubes were smoothed by a $15''$--FWHM 1D--Gaussian
      along the $y$ direction perpendicular to the illuminating star
      direction. Due to their small field of view (in particular in the $y$
      direction), the \twCO{} \Jtwo{} data were just smoothed by a
      $5''$--FWHM circular Gaussian.}
    \label{fig:cuts:spectra} 
  \end{figure*}}

\newcommand{\FigObsPicoVeleta}{%
  \begin{figure*}[ht]
    \centering %
    \includegraphics[width=\hsize{}]{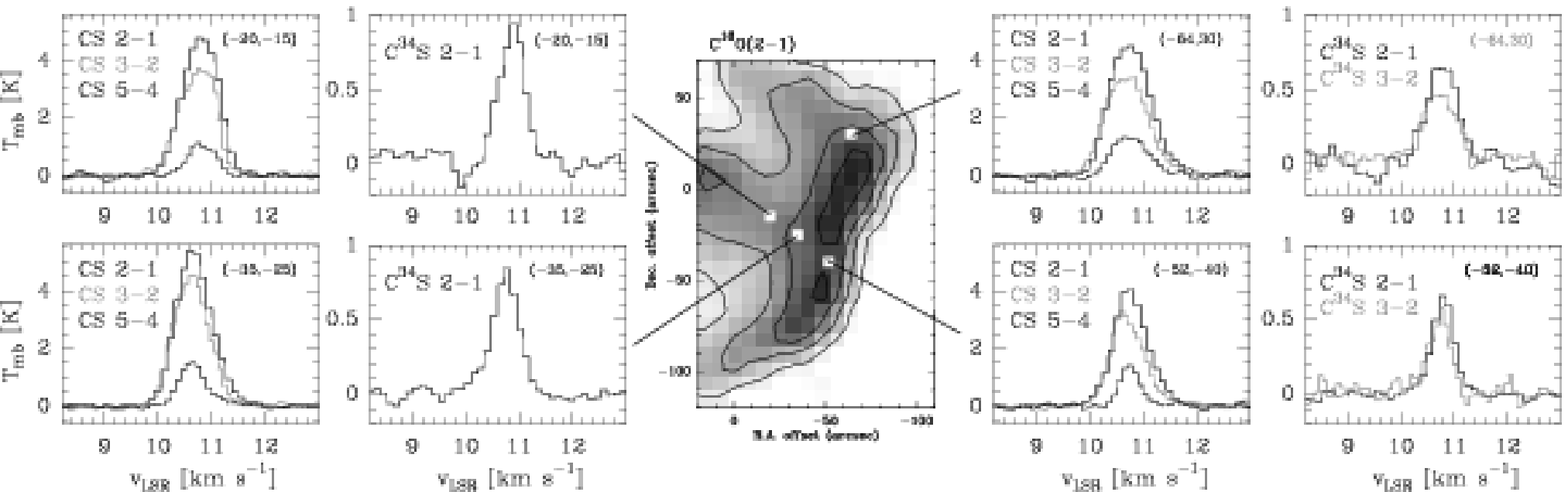}
    \caption{IRAM--30m CS $J$=2--1, 3--2 and 5--4, and C$^{34}$S $J$=2--1 and
    3--2 single--dish observations (histograms) at different positions
    of the Horsehead PDR single--dish C$^{18}$O $J$=2--1 emission 
    centered at $\alpha_{2000} = 05^h40^m58^s $, $\delta_{2000} =
        -02^\circ 27' 20''$ (from Hily--Blant et al. 2005).} 
    \label{fig:obs_picoveleta} 
  \end{figure*}}

\newcommand{\FigModsCS}{%
  \begin{figure*}[hb]
    \centering %
    \includegraphics[width=\hsize{},angle=0]{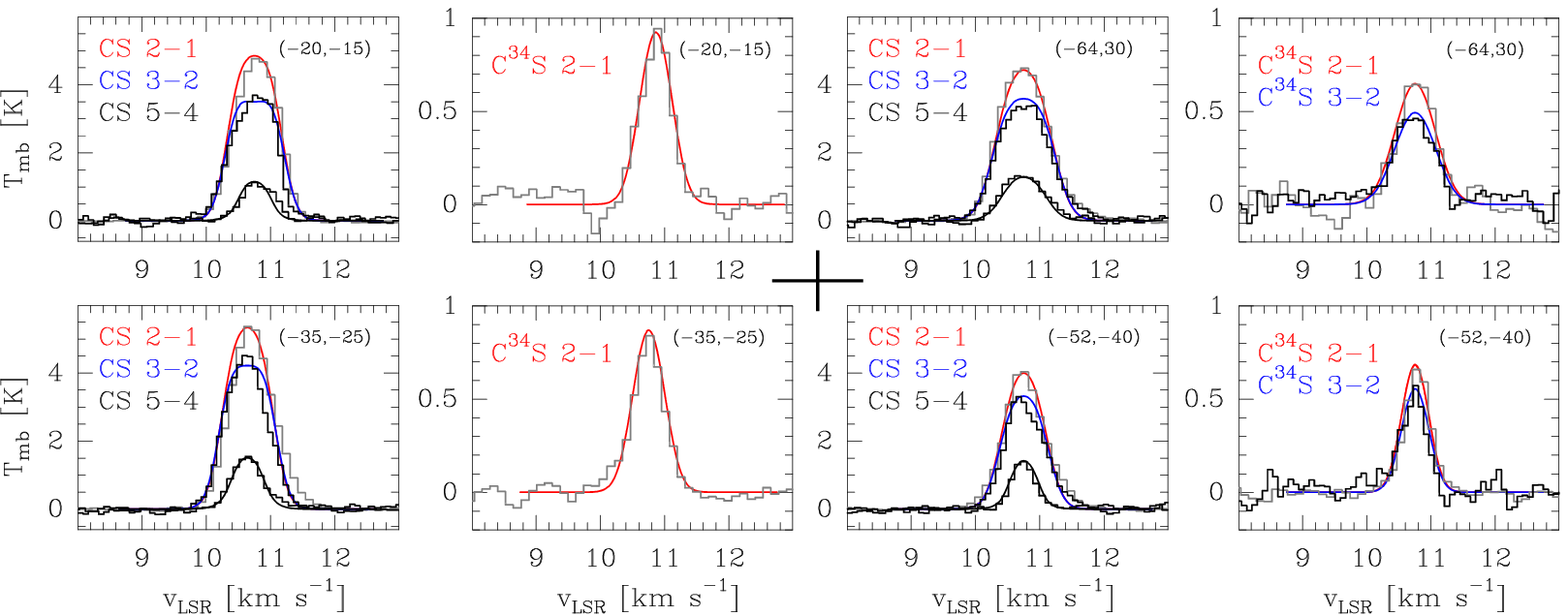}
    \caption{Radiative transfer models for CS and C$^{34}$S discussed in the text  (curves)
    that best fits the IRAM--30m observations (histograms).
    Offsets in arcsec refer to the (0,0) position of the C$^{18}$O(2--1) map
    (see Fig.\ref{fig:obs_picoveleta}).  Predicted line profiles have been convolved 
    with the telescope angular resolution at each frequency.
    Intensity scale is in main beam temperature.}
    \label{fig:mods_cs}
  \end{figure*}}

\newcommand{\FigObsHCSP}{%
  \begin{figure*}[hbt]
    \centering %
    \includegraphics[width=4.5cm,angle=-90]{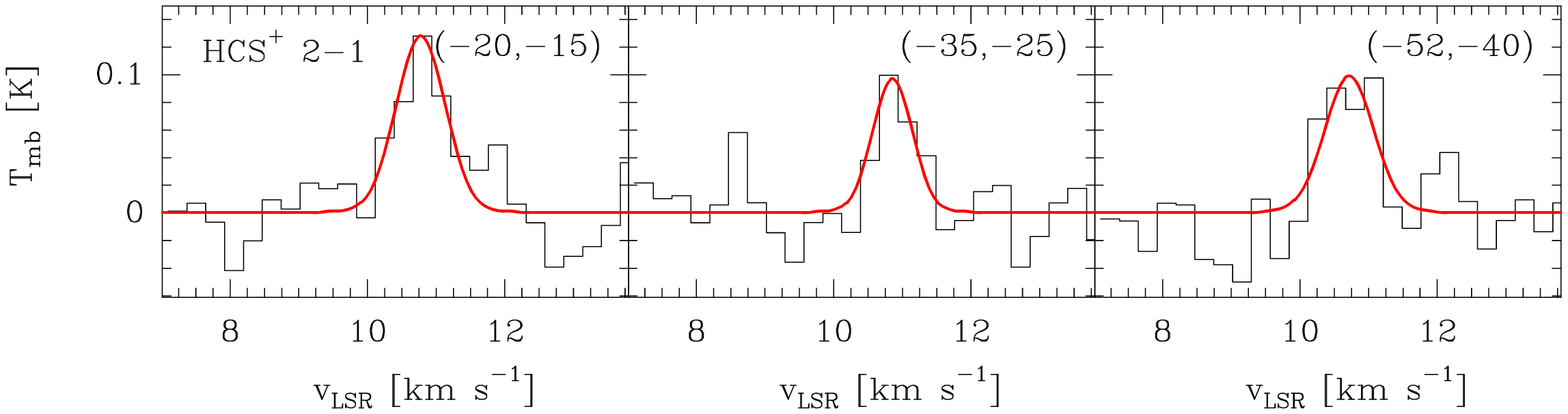}
    \caption{IRAM--30m HCS$^+$(2--1) single--dish observations 
    (histograms) at different positions of the Horsehead.
    Offsets in arcsec refer to the (0,0) position of the C$^{18}$O(2--1) map
    (see Fig.\ref{fig:obs_picoveleta})  Radiative transfer models for HCS$^+$ at selected
    positions are also shown (curves). 
    Predicted line profiles have been convolved with the telescope 
    angular resolution at each frequency.
    Intensity scale is in main beam temperature.}
    \label{fig:obs_hcsp} 
  \end{figure*}}

\newcommand{\FigRotdiag}{%
  \begin{figure}[b]
    \centering %
    \includegraphics[width=6.1cm,angle=-90]{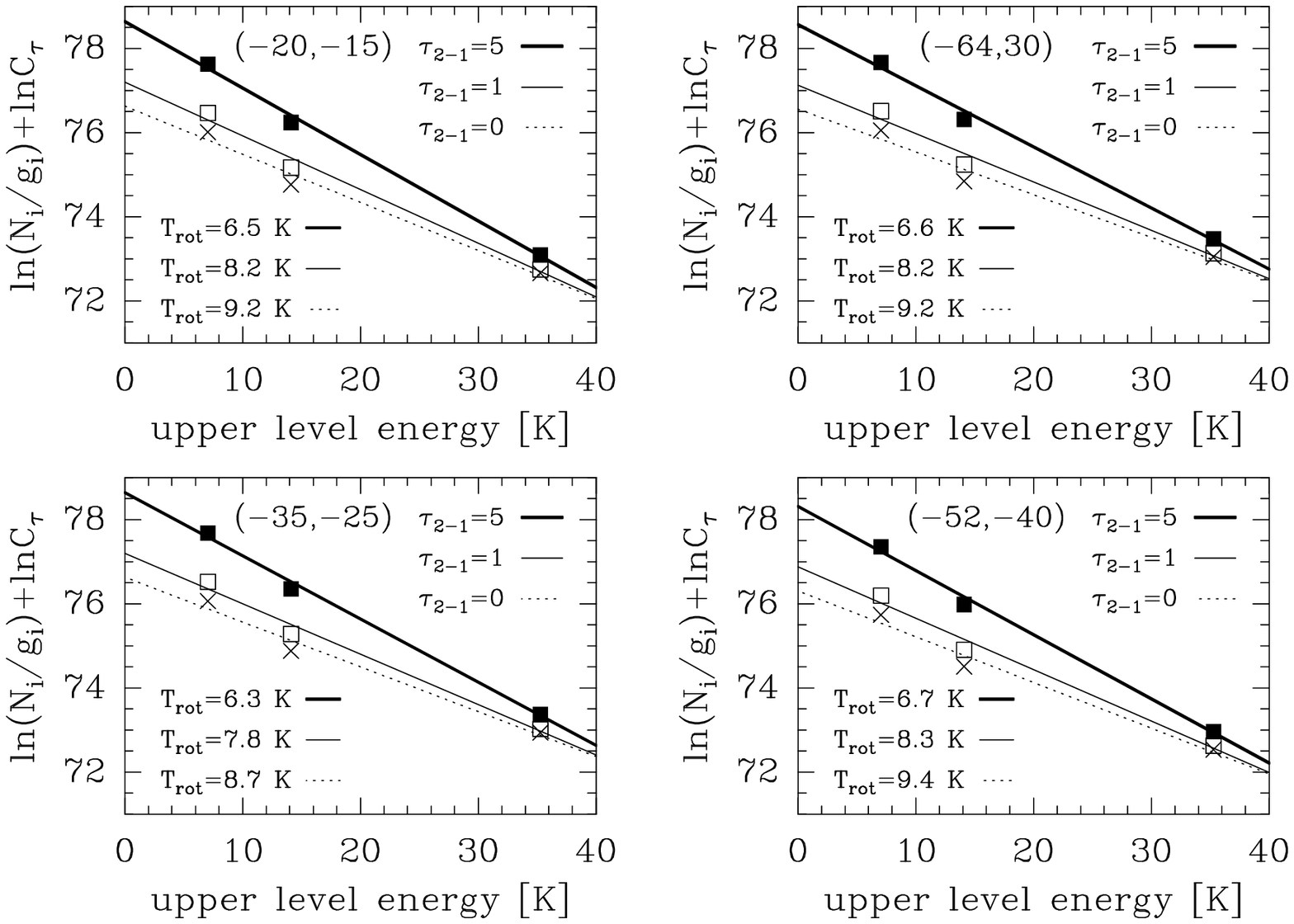}
    \caption{CS rotational--diagrams corrected for line opacity 
    effects at each observed position of Fig.\ref{fig:obs_picoveleta}. 
    Rotational--diagrams for different considered CS $J$=2--1 
    line opacities ($\tau_{2-1}$) are shown in each box. Rotation temperatures for
    each opacity correction are also indicated.}
    \label{fig:ROTdiag} 
  \end{figure}}

\newcommand{\FigMTCgridCS}{%
  \begin{figure*}[h]
    \centering %
    \includegraphics[width=4.2cm,angle=-90]{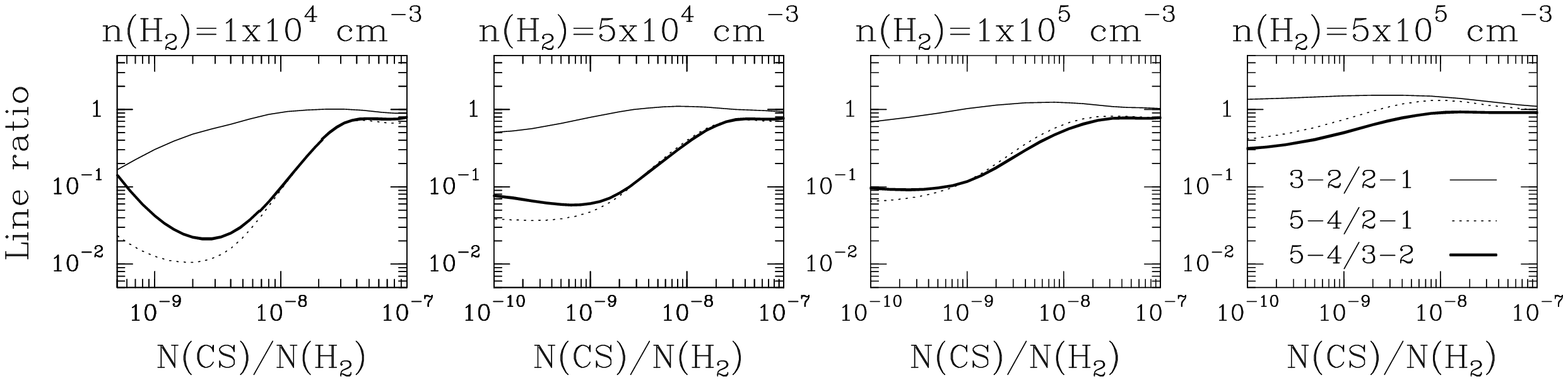}
    \caption{Grid of CS single--component models 
    assuming  T$_k$= 30~K  and a fixed extinction of 20 mag.
    Panels show different line  ratios as a function of $\chi(CS)$. 
    Each panel correspond to a single density, from $n$(H$_2$)=10$^4$
    to 5$\times$10$^5$~cm$^{-3}$. Mean observed ratios are 
    $\frac{3-2}{2-1}$$\simeq$0.7, $\frac{5-4}{2-1}$$\simeq$0.2 
    and $\frac{5-4}{3-2}\simeq$0.3.}
    \label{fig:MTCgrid} 
  \end{figure*}}

\newcommand{\FigPdrCSHCSP}{%
  \begin{figure}[th]
    \centering %
    \includegraphics[width=6.cm, angle=-90]{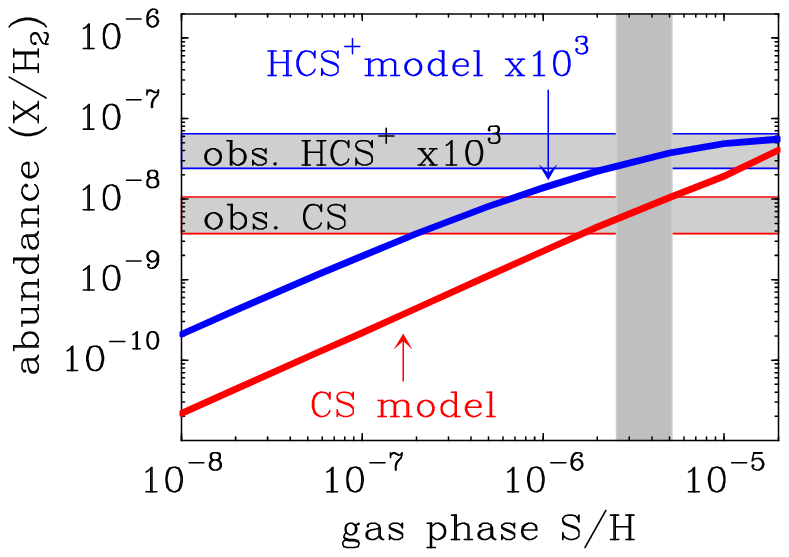}
    \caption{Photochemical model predictions for the physical and 
    FUV illuminating conditions prevailing in the Horsehead PDR showing 
    the CS and HCS$^+$ abundance as a function of the sulfur gas phase 
    abundance. Horizontal shaded regions show the CS and HCS$^+$ abundances 
    derived from the single--dish observations and radiative transfer modeling.
    Note that for clarity HCS$^+$ abundances have been multiplied by a 
    factor of 1000. The shaded vertical region shows the estimated sulfur 
    abundance in the  Horsehead nebula derived from the constrained fits
    of CS and HCS$^+$ abundances.}
    \label{fig:cs_hcsp_pdr}
  \end{figure}}

\newcommand{\FigModelgeo}{%
  \begin{figure}[b]
    \centering %
    \includegraphics[width=7.5cm]{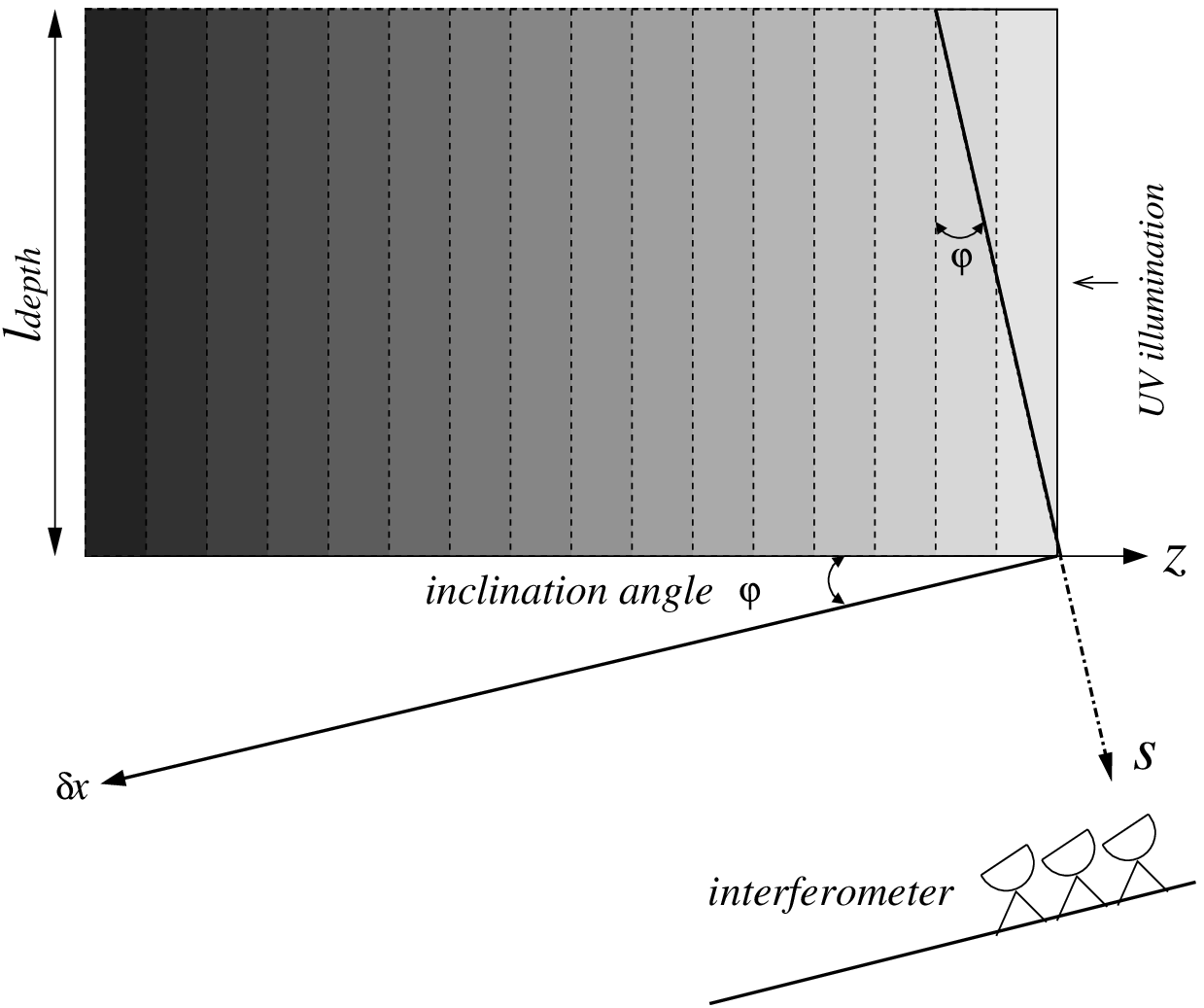}
    \caption{Adopted geometry for a plane-parallel PDR inclined by an
    small angle $\varphi$ relative to the line of sight $s$. 
    In this sketch, $z$ denotes the normal direction to the slabs 
    and also the UV illumination direction.}
    \label{fig:geometry}
  \end{figure}}

\newcommand{\FigModsPDR}{%
  \begin{figure*}
    \centering %
    \includegraphics[width=14cm,angle=-90]{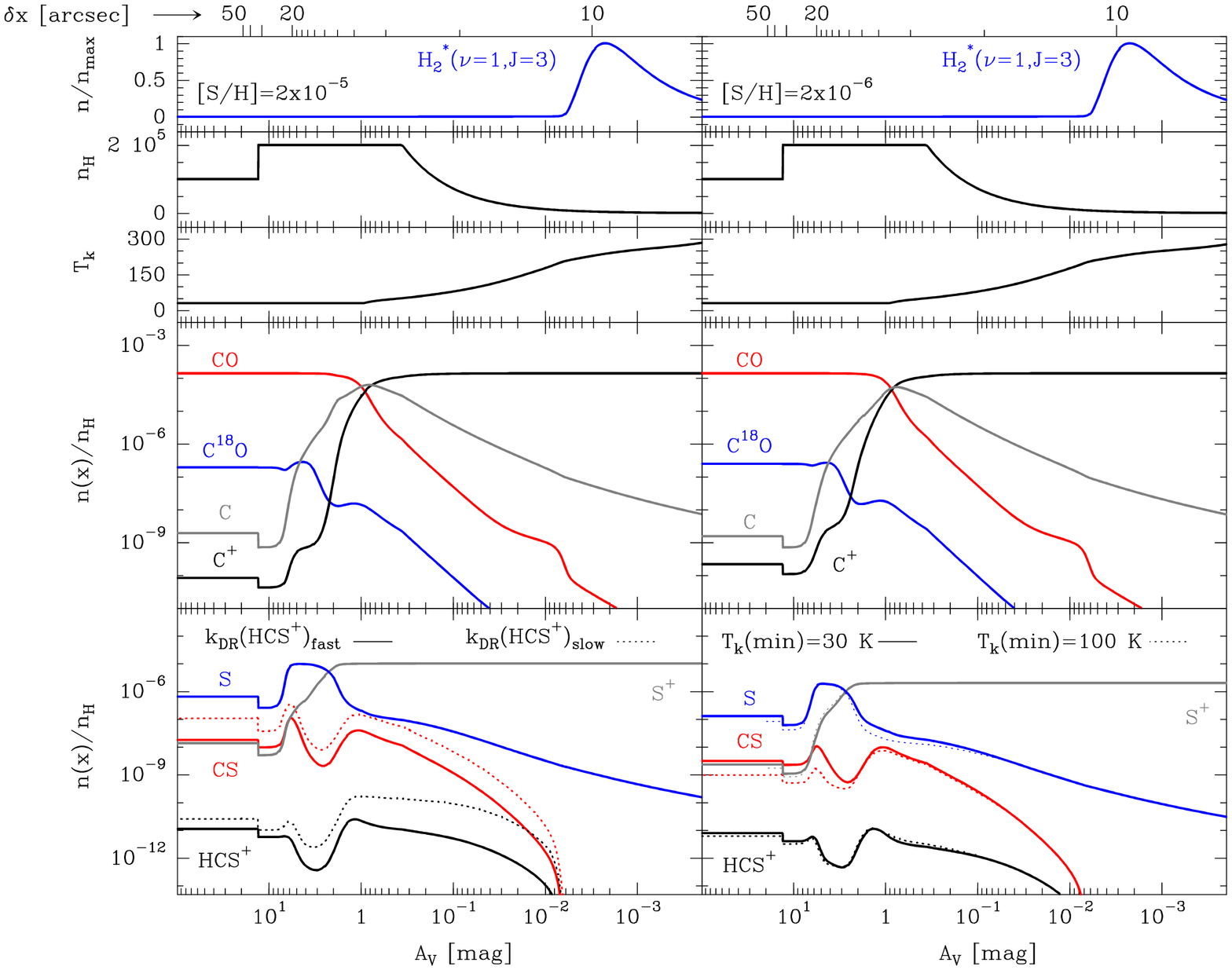}
    \caption{Photochemical models using a unidimensional PDR code
    for two different sulfur gas phase abundances,  (S/H=2$\times$10$^{-5}$; $left$) and 
    the minimum value found for the Horsehead (S/H=2$\times$10$^{-6}$; $right$). 
    The predicted normalized population  of the H$_2$ $v$=1, $J$=3 level is shown in 
    the $upper$ panel and is used to place the $\delta$x--axis origin for the models. 
    The peak of this curve is placed at the maximum of the observed H$_{2}$ 1--0 S(1)
    2.12~$\mu$m line emission  ($\delta$x$\sim$10$''$;  Habart et al. 2005). Next panel shows
    the density profile  ($n_H = n(H)+2n(H_2)$ in cm$^{-3}$) used in the PDR calculations 
    that better fits the CS and C$^{18}$O  IRAM-PdBI observations.
    Next panel shows the gas temperature (in K) consistently computed in thermal balance 
    until reaches a minimum value of 30~K.
    $Lower$ panels show the spatial variation of C$^{18}$O/CO/C/C$^+$ and
    CS/HCS$^+$/S/S$^+$ abundances (relative to $n_H$) across the PDR. The far--UV radiation
    field is $\chi$=60 times the Draine field. Chemical rates are those 
    of the \textit{Ohio State University (osu)}  gas--phase chemical network (September 
    2005 release) plus several modifications (see text).
    \textit{Bottom left} panel shows the effect of using the older rate and
    branching ratios for the HCS$^+$ dissociative recombination 
    on the CS and HCS$^+$ abundances (dashed curves).
    \textit{Bottom right} panel shows the effect of using a minimum gas
    temperature of 100~K in the chemistry. Lower CS abundances
    and thus larger S/H values are possible as the temperature increases (dashed curves).}
  \label{fig:ModsPDR} 
  \end{figure*}}

\newcommand{\FigPdBIMTC}{%
  \begin{figure*}
    \centering %
    \includegraphics[width=6cm,angle=-90]{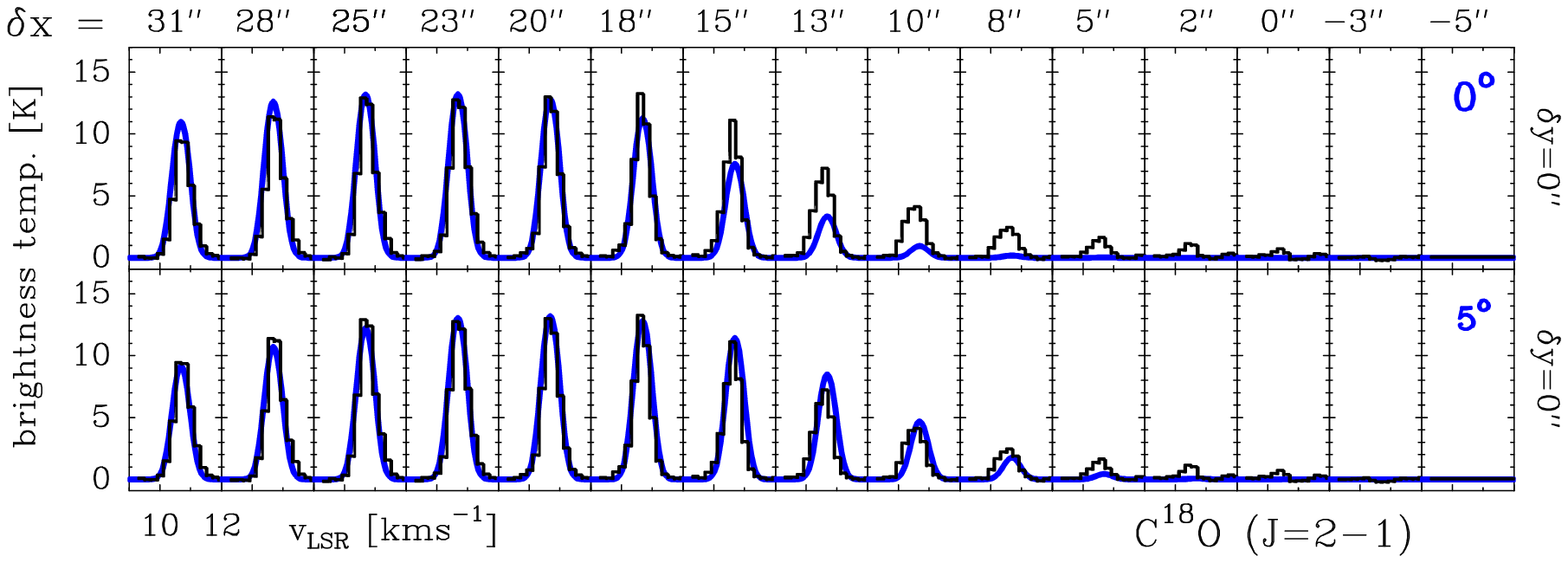}
    \caption{IRAM-PdBI C$^{18}$O $J$=2--1 spectra along
    the direction of the exciting star at $\delta y = 0''$ (histograms). 
    Radiative transfer models using the output of PDR models 
    for C$^{18}$O (blue curve)  
    for a density gradient and physical conditions discussed in the text. 
    \textit{Lower panel} shows inclination effects assuming that   the PDR
    is inclined relative to the line of sight by a $\varphi$=5$^o$ angle.
    Modeled line profiles have been convolved with an appropriate gaussian
    beam corresponding to each synthesized beam.  
    Intensity scale is in brightness temperature and abscissa in LSR velocity.}
    \label{fig:PdbImtc} 
  \end{figure*}}

	\newcommand{\FigCSPdBIMTCa}{%
  \begin{figure*}
    \centering %
    \includegraphics[width=6cm,angle=-90]{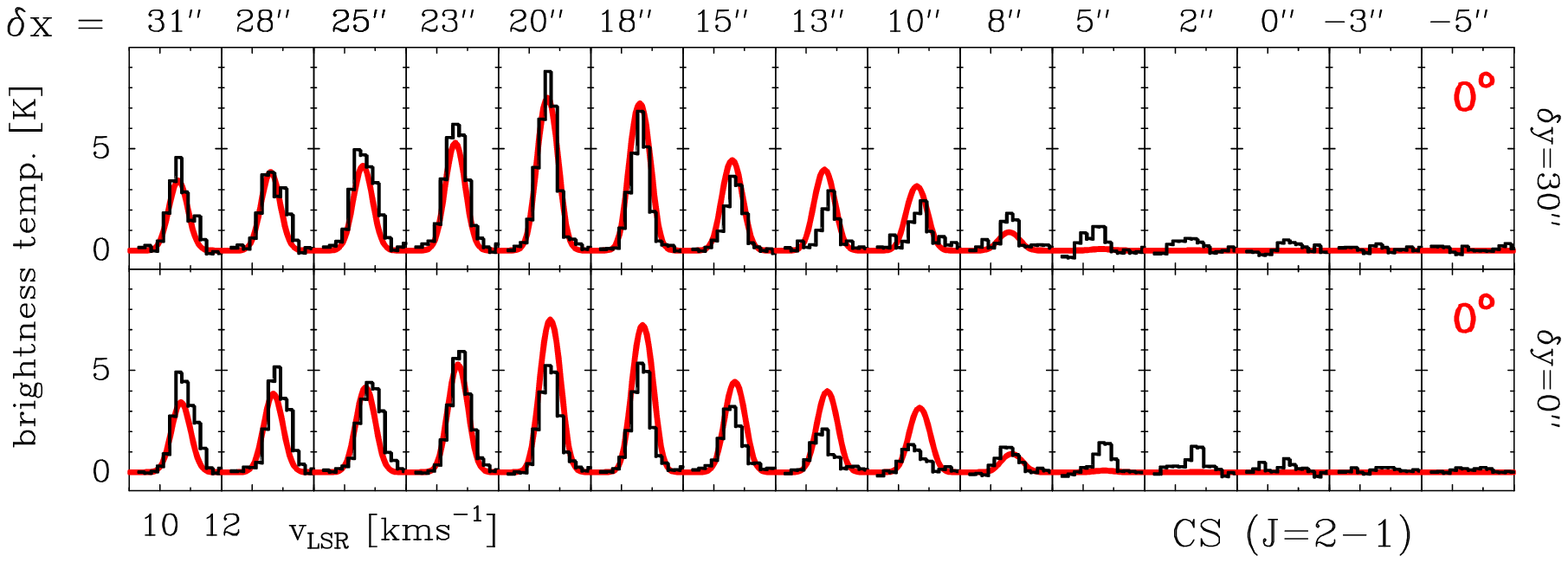}
    \caption{IRAM-PdBI CS $J$=2--1 spectra along
    the direction of the exciting star at $\delta y = 30''$ ($upper panel$)
    and $\delta y = 0''$ ($lower panel)$. 
    Radiative transfer models using the output of PDR models 
    for CS (red curve) for a density gradient and physical conditions 
    discussed in the text (assuming that the PDR
    is not inclined relative to the line of sight).
    Modeled line profiles have been convolved with an appropriate gaussian
    beam corresponding to each synthesized beam.  
    Intensity scale is in brightness temperature and abscissa in LSR velocity.} 
    \label{fig:CSPdbImtca} 
  \end{figure*}}

\newcommand{\FigCSPdBIMTCb}{%
  \begin{figure*}
    \centering %
    \includegraphics[width=6cm,angle=-90]{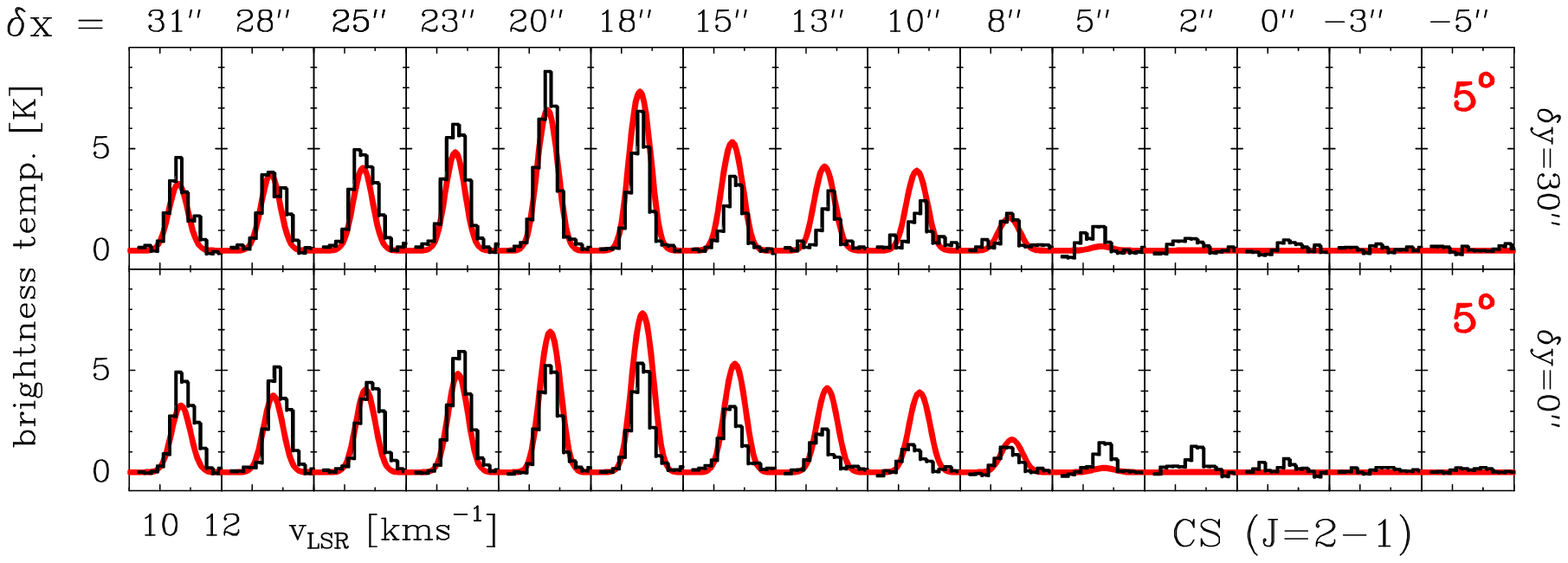}
    \caption{IRAM-PdBI CS $J$=2--1 spectra along
    the direction of the exciting star at $\delta y = 30''$ ($upper panel$)
    and $\delta y = 0''$ ($lower panel$). 
    Radiative transfer models using the output of PDR models 
    for CS (red curve) for a density gradient and physical conditions 
    discussed in the text (assuming that the PDR
    is inclined relative to the line of sight by a $\varphi$=5$^o$ angle).
    Modeled line profiles have been convolved with an appropriate gaussian
    beam corresponding to each synthesized beam.  
    Intensity scale is in brightness temperature and abscissa in LSR velocity.} 
    \label{fig:CSPdbImtcb} 
  \end{figure*}}

\newcommand{\FigHalos}{%
  \begin{figure}
    \centering %
    \includegraphics[width=7.5cm, angle=-90]{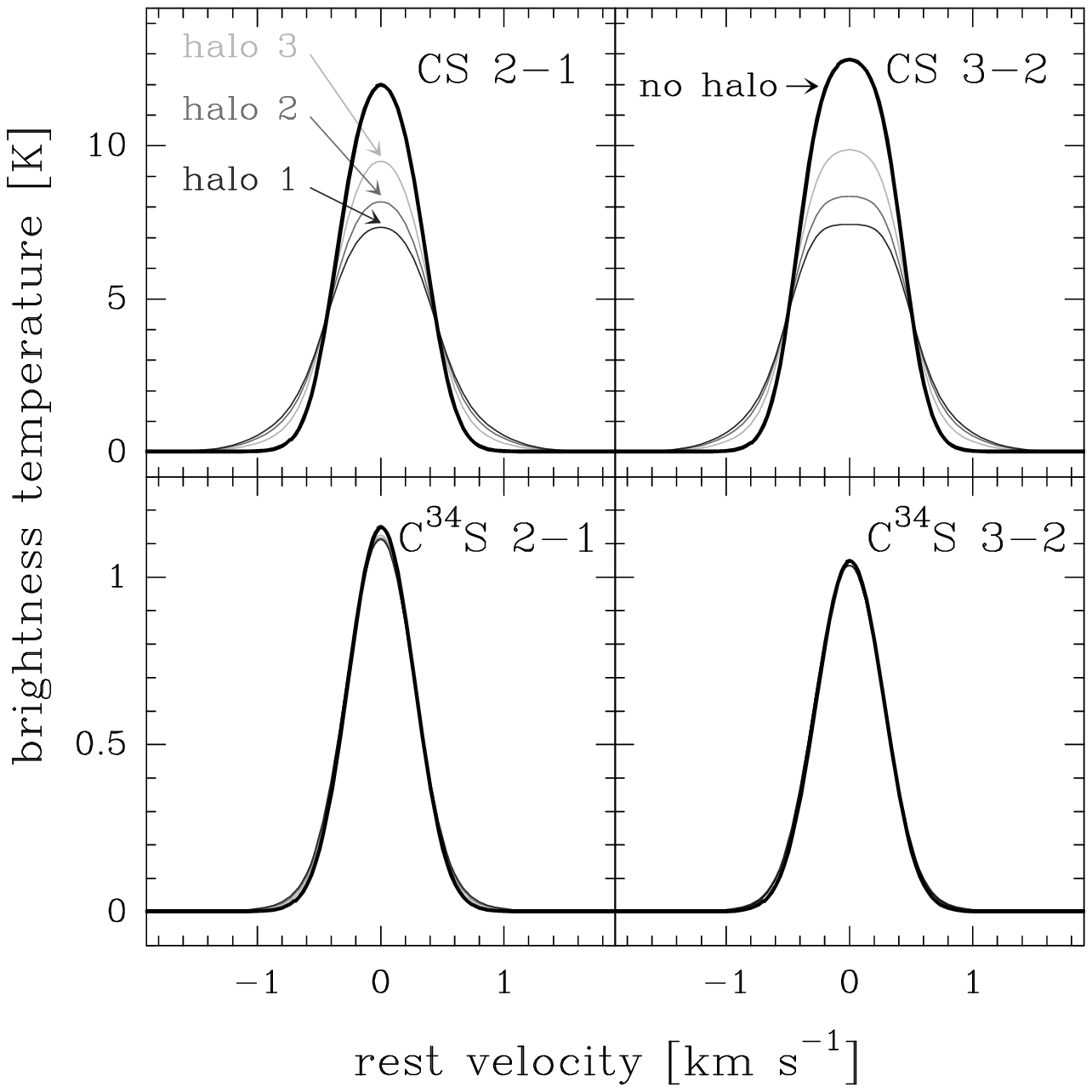}
    \caption{CS and C$^{34}$S synthetic line profiles for a 
    cloud with a depth of 0.1 pc, T$_k$=30~K, $n$(H$_2$)=10$^5$~cm$^{-3}$ 
    and $\chi$(CS)=7$\times$10$^{-9}$ (thick curves). Thin curves show 
    the resulting spectra if the same cloud is surrounded by different 
    diffuse halos (3: $n$(H$_2$)=5$\times$10$^3$~cm$^{-3}$,
    2: $n$(H$_2$)=8$\times$10$^3$~cm$^{-3}$ and
    1: $n$(H$_2$)=1$\times$10$^4$~cm$^{-3}$). The
    CS abundance in the cloud is determined more precisely from 
    CS high-$J$ and  C$^{34}$S low--$J$ observations, otherwise it is
     underestimated. Note that the
    intensity levels are comparable to those observed in the Horsehead.}
    \label{fig:Halos}
  \end{figure}}

\newcommand{\FigPlgeo}{%
  \begin{figure}[t]
    \centering %
    \includegraphics[width=8.5cm]{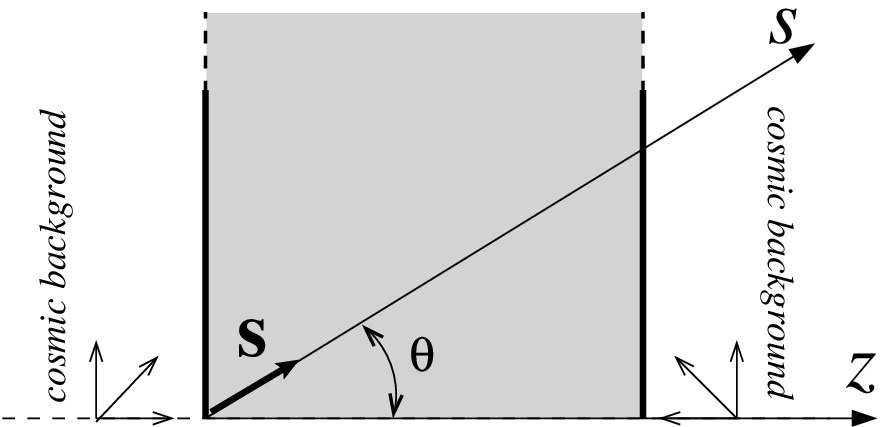}
    \caption{Plane--parallel geometry for a cloud isotropically
    illuminated by the cosmic microwave background at both surfaces.}
    \label{fig:_plgeo}
  \end{figure}}

\newcommand{\FigCSbench}{%
  \begin{figure*}[ht]
    \centering %
    \includegraphics[width=4.5cm, angle=-90]{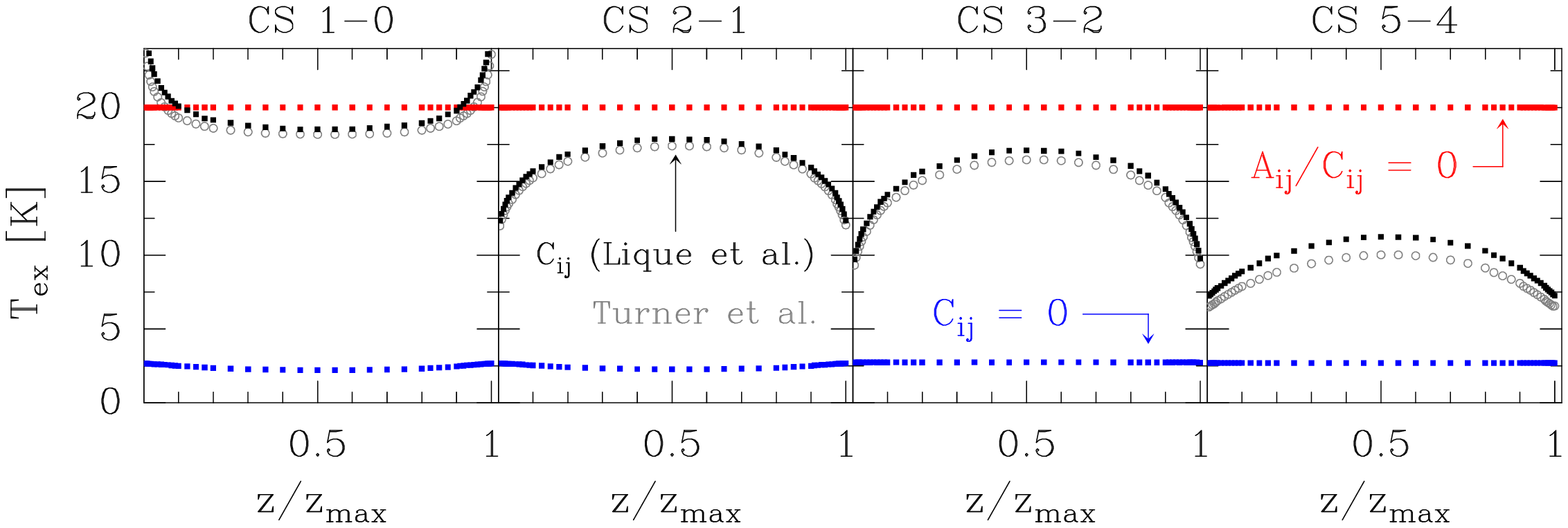}
    \caption{Thermalization tests for a plane-parallel cloud illuminated
    by the cosmic background at both surfaces ($z/z_{max}$=0,1).} 
    \label{fig:CS_bench}
  \end{figure*}}

\begin{document}
\title{Low sulfur depletion in 
the Horsehead PDR\thanks{Based on observations obtained 
with the IRAM Plateau de Bure interferometer 
and 30~m telescope. IRAM is supported by INSU/CNRS (France), MPG (Germany), and IGN (Spain).}}

%\subtitle{I. Overviewing the $\kappa$-mechanism}

\author{J.R. Goicoechea\inst{1}%
\and J. Pety \inst{1,2}%
\and M. Gerin \inst{1}%
\and D. Teyssier \inst{3}%
\and E. Roueff \inst{4}%
\and P. Hily-Blant \inst{2}%
\and S. Baek \inst{1}}

\offprints{\email{javier@lra.ens.fr}}

\institute{
LERMA--LRA, UMR 8112, CNRS, Observatoire de Paris and Ecole Normale
Sup\'erieure, 24 Rue Lhomond, 75231 Paris cedex 05, France.\\
\email{javier@lra.ens.fr, gerin@lra.ens.fr}%
\and{} IRAM, 300 rue de la Piscine, 38406 Grenoble cedex, France.\\
\email{pety@iram.fr,hilyblan@iram.fr}
\and European Space Astronomy Centre, Urb. Villafranca del Castillo, P.O.
Box 50727, Madrid 28080, Spain.\\
\email{dteyssier@sciops.esa.int}
\and LUTH UMR 8102, CNRS and Observatoire de Paris, Place J. Janssen
92195 Meudon cedex, France.\\
\email{evelyne.roueff@obspm.fr}}

\date{Received 2006 March 24; accepted 2006 June 9}

% \abstract{}{}{}{}{} 
% 5 {} token are mandatory
 
 \abstract
% context heading (optional)
%{CS is one of the most widely used diagnostics of dense gas in star forming regions.
%However, sulfur chemistry is far for being fully understood. Recent improvements in CS collisional %and chemical rates may help to make progresses in our understanding of sulfur chemistry.}
% aims heading (mandatory)
{}
{We present  $3.65'' \times 3.34''$ angular--resolution IRAM \textit{Plateau de Bure 
Interferometer} (PdBI)  observations of the CS $J$=2--1  line toward the 
Horsehead \textit{Photodissociation Region} (PDR), complemented with IRAM--30m 
single--dish observations of several rotational lines of CS, C$^{34}$S and HCS$^+$. 
We analyse the CS and HCS$^+$  photochemistry, excitation and radiative transfer 
to obtain their abundances and the physical conditions prevailing in the cloud edge.
Since the CS abundance scales to that of sulfur, we  determine the gas phase sulfur 
abundance in the PDR, an interesting intermediate medium  between translucent clouds
 (where sulfur remains in the gas phase) and dark clouds 
(where large depletions have been invoked).}
% methods heading (mandatory)
{A nonlocal non--LTE radiative transfer code including dust and cosmic background 
illumination adapted to the Horsehead  geometry has been developed to carefuly analyse
the  CS, C$^{34}$S, HCS$^+$ and C$^{18}$O rotational line emission. 
We use this model to consistently link the line  observations with photochemical 
models  to determine the CS/HCS$^+$/S/S$^+$ structure of the PDR.}
% results heading (mandatory)
{Densities of $n(H_2)\simeq(0.5-1.0)\times$10$^5$~cm$^{-3}$ are required to reproduce 
the  CS and C$^{34}$S $J$=2--1 and 3--2 line emission. CS $J$=5--4 lines show narrower line widths than 
the  CS low--$J$  lines  and  require higher density gas components not  
resolved by the  $\sim$10$''$ IRAM--30m beam.
These values are larger than previous estimates based in CO observations. 
We found $\chi(CS)$=(7$\pm$3)$\times$10$^{-9}$ and
$\chi(HCS^{+})$=(4$\pm$2)$\times$10$^{-11}$ as the averaged abundances in the PDR.
According to photochemical models, the gas phase sulfur abundance  required to reproduce
 these values is  S/H=(3.5$\pm$1.5)$\times$10$^{-6}$, only a factor $\lesssim$4 less 
abundant than the solar sulfur elemental abundance. Since only lower limits to the 
gas temperature are constrained, even lower sulfur depletion values are possible if 
the gas is significantly warmer.}
{The combination of  CS, C$^{34}$S and HCS$^+$ observations together with the inclusion
of the most recent CS collisional  and chemical rates in our 
models implies that sulfur depletion invoked 
to account for CS and HCS$^+$ abundances is much smaller than in previous studies.}

% javi

\maketitle
%
%________________________________________________________________

\section{Introduction}
\label{sec:introduction}

Sulfur is an abundant element (the solar photosphere abundance is 
S/H=1.38$\times$10$^{-5}$; \cite{asp05}), which remains undepleted in  diffuse interstellar 
gas (e.g. \cite{how06}) and HII regions (e.g. \cite{mar02}; \cite{gar06} and references
therein) but it is historically assumed
to deplete on grains in higher density molecular clouds by factors as large as $\sim$10$^3$ 
(Tieftrunk et al. 1994). This conclusion is simply reached by adding up the observed gas 
phase abundances of S--bearing molecules in well known dark clouds such as TMC1 
(e.g. Irvine et al. 1985; Ohishi \& Kaifu 1998).  As sulfur is easily ionized 
(ionization potential $\sim$10.36~eV), 
sulfur ions are probably the dominant gas--phase sulfur species in translucent gas. 
Ruffle et al. (1999) proposed that if dust grains are typically negatively charged, S$^+$ may 
freeze--out onto dust grains during cloud collapse more efficiently than neutral species
such as oxygen. However, the nature of sulphur on dust  grains (either in mantles or 
cores) is not obvious. Van der Tak et al. (2003) observed large abundances of gas phase
OCS, $\sim$10$^{-8}$,  in star forming regions, and suggested that together with  the detection
 of solid OCS (with an abundance of $\sim$10$^{-7}$; Palumbo et al. 1997), it
implies that OCS is a major sulfur carrier in dust grains. However, the $\sim$4.9~$\mu$m ice 
feature attributed to OCS is best reproduced when OCS is mixed with methanol. In fact,
the band is blended with a methanol overtone whose contribution has not been studied in 
detail (Dartois 2005). In any case, the absence of strong IR features due to S--bearing
ices in many ISO's mid--IR spectra (e.g. Boogert et al. 2000; Gibb et al. 2004) and the 
presence of S\,{\sc ii} recombination lines in dark clouds such as Rho Ophiuchi 
(Pankonin \& Walmsley 1978) all argue against a large depletion of sulfur from the gas phase.
In this case, the abundance of species such as CS may indicate that something 
important is lacking from chemical models or that an abundant sulfur--bearing carrier
has been missed.  Therefore, the abundances of sulfur species remain interesting puzzles 
for interstellar chemistry. In the case of dense clouds, standard chemical models 
predict that most of the gas phase sulfur is shared between S, SO and CS 
(Millar \& Herbst 1990), while H$_2$S 
is also abundant in the  Orion Bar PDR (Jansen et al. 1995). In all these cases, a large 
sulfur depletion, $\sim$10$^2$,  was required in the models to explain the observed 
abundances.

PDRs offer an ideal intermediate medium between diffuse and dark cloud gas to investigate 
the sulfur depletion problem. In this work we have tried to  determine the CS abundance in
the Horsehead PDR as a tool for estimating  the sulfur gas phase abundance. However, CS 
chemistry is an open issue itself in different environments, from hot cores
(e.g.~\cite{wak04}) to extragalactic sources (e.g.~\cite{mar05}).  
Recent  laboratory experiments on dissociative 
recombination of  HCS$^+$ and OCS$^+$ (Montaigne et al. 2005)  imply a substantial 
modification of previous reaction rate coefficients, dissociative channels and branching
 ratios used in chemical models. The latest available reaction rates and collisional 
coefficients have been used in our photochemical and radiative transfer models.

\subsection{The Horsehead nebula}

The Horsehead nebula, appears as a dark patch of $\sim$5$'$ diameter against the bright HII
 region IC434. Emission from  gas and dust associated with this globule has been detected 
from IR to millimeter wavelengths (\cite{ab02,ab03,prb03,tf04}, Habart et al. 2005, 
Pety et al. 2005a),  although the first astronomical plates were taken $\sim$120~yr ago.  
In particular, the Horsehead western edge is a PDR viewed nearly edge-on and illuminated by 
the O9.5V star $\sigma$Ori  at a projected distance of $\sim$3.5\pc{} (\cite{at82}). 
The intensity  of the incident FUV radiation field is $\chi$ $\simeq$60 relative to the 
interstellar radiation field (ISRF) in Draine's units (\cite{dra78}). 

According to the evolutionary view of Reipurth \& Bouchet (1984), the Horsehead nebula was
 a quiescent and dense cloud core embedded in a more diffuse cloud (L1630). 
The erosive action of the UV radiation from  $\sigma$Ori on the ambient gas led to the
 apparent emergence of the core cloud, as in the earliest stages of Bok globules still 
attached 
to their parental cloud. However, the observed morphology together with  the velocity 
gradients of the cloud, require a more involved description including a pre--existing 
rotating velocity field as well as  density inhomogeneities in the initial structures 
(Pound et al. 2003; Hily--Blant et al. 2005). The erosive effect of the ionizing and 
dissociating radiation field  together with these initial conditions explain the peculiar 
shaping of the Horsehead nebula. 
In particular, the densest regions of the initial inhomogeneities are now believed to be
 the East-West filamentary material  connecting it to the parental cloud, and the PDR.
In this work we have studied the PDR through CS, C$^{34}$S and HCS$^+$ observations.

\FigMaps{} 

\TabObs{} %

\section{Observations and data reduction}
\label{sec:obs}

\subsection{Observations}

\subsubsection{Pico Veleta single--dish}

The single-dish data presented in this paper have been gathered between February and 
October 2004 at the IRAM 30-m telescope. The Horsehead nebula PDR was mapped in the 
CS $J$=2--1 and 5--4 lines in order to provide the short-spacings for the interferometric
 observations presented thereafter. The final map consists of 5 on-the-fly coverages performed 
along perpendicular scanning directions, and combined with the PLAIT algorithm introduced 
by Emerson \& Gr{\"a}ve (1988), allowing to efficiently reduce the stripes over the map. 
The noise levels (1$\sigma$ rms) per regridded pixel and resolution channel of 80 kHz are 
of the order of  0.15 K at 3\,mm, and 0.64\,K at 1.3\,mm. The latter value  was not low 
enough to provide any useful mapping information at 1.3\,mm since the CS $J$=5--4
line peak is  $\lesssim$1\,K.

In complement to these data, dedicated positions were probed over a larger set of species 
and transitions. The frequency switching mode was used to observe CS $J$=2--1, 3--2 and
5--4~lines, as well as C$^{\rm 34}$S $J$=2--1, 3--2, and HCS$^+$ $J$=2--1 lines. 
Table~\ref{cs param} summarizes the corresponding observing parameters. Longer integrations 
allowed to reach, in a resolution channel of 40 kHz, rms noise levels of 25, 42 and 36\,mK 
at 3, 2, and 1.3\,mm respectively. All  CS and C$^{34}$S lines were detected with 
a S/N ratio better than 10. 
Figs.~\ref{fig:obs_picoveleta}  and \ref{fig:obs_hcsp} show some spectra collected at 
positions  inside and across the PDR.

%%%%%%%%%%%%%%%%%%%%%%%%%%%%%%%%%%%%%%%%%%%%%%%%%%%%%%%%
\begin{table}
\caption{Line parameters for the IRAM 30-m CS observations.}
\begin{center}
\begin{tabular}{l c c c } 
\hline \hline  
Molecule   & Transition	&Frequency	&HPBW	  \\
           &            &(GHz)  	&(arcsec) \\ \hline \hline
CS	   & $J$=2--1 	& 97.980968     & 25	  \\
           & $J$=3--2 	& 146.96905 	& 16   	  \\
           & $J$=5--4 	& 244.93561 	& 10   	  \\
C$^{34}$S  & $J$=2--1 	& 96.412982     & 25   	  \\
           & $J$=3--2 	& 144.61715 	& 16 	  \\
HCS$^+$    & $J$=2--1 	& 85.347884     & 29   	  \\		
\hline
\end{tabular}
\label{cs param}
\end{center} 
\end{table} 

%%%%%%%%%%%%%%%%%%%%%%%%%%%%%%%%%%%%%%%%%%%%%%%%%%%%%%%%

The data were first calibrated to the $T_{\rm A}^*$ scale using the
so-called chopper wheel method (Penzias \& Burrus 1973), and finally
converted to main beam temperatures using 
efficiencies (B$_{\rm eff}$/F$_{\rm eff}$)  of 0.81, 0.74 and 0.50 at 3, 2
and 1.3\,mm respectively.

\subsubsection{Plateau de Bure Interferometer}

\PdBI{} observations dedicated to this project were carried out with 6
antennae in the BCD configuration (baseline lengths from 24 to 331~m) from
August 2004 to March 2005. The 580\MHz{} instantaneous IF--bandwidth
allowed us to simultaneously observe CS, \lCCCH{} and \tfSO{} at
3\mm{} using 3 different 20\MHz{}--wide correlator windows. Another
window was centered on the \thCO{} \Jtwo{} line frequency at 1\mm{}. The full IF
bandwidth was also covered by continuum windows both at 3.4 and 1.4\mm{}.
Only CS \Jtwo{} and \thCO{} \Jtwo{} (not shown here) were detected.

We observed a seven--field mosaic. The mosaic was Nyquist sampled in
declination at 3.4~mm and Nyquist sampled in Right Ascension at 1.3~mm.
This ensures correct sampling in the illuminating star direction both at 3
and 1~mm while maximizing the field of view along the edge of the PDR
eases the deconvolution. 
This mosaic, centered on the IR peak (Abergel et al. 2003), 
was observed for about 30~hours of \emph{telescope} time with 6 antennas.
This leads to an \emph{on--source} integration time of useful data of 10~hours
after filtering out data with tracking errors larger than $1''$ and with
phase noise worse than 40$\deg$ at 3.4\mm{}.  The rms phase noises were
between 15 and 40\deg{} at 3.4\mm{}, which introduced position errors $<
0.5''$.  Typical 3.4\mm{} raw resolution was $2.5''$.

\TabFlux{} %

\subsection{Data processing}

All data reduction was done with the \GILDAS{}\footnote{See
\texttt{http://www.iram.fr/IRAMFR/GILDAS} for more information about the
 \GILDAS{} software.} softwares supported at IRAM (Pety 2005b).
Standard calibration methods using nearby calibrators were applied to all
the \PdBI{} data. The calibrator fluxes used for the absolute flux
calibration are summarized in Table~\ref{tab:fluxes}.

Following~\cite{gu96}, single--dish, fully sampled maps obtained with the
IRAM-30m telescope (see section~2.1.1) were used to produce the short--spacing
visibilities filtered out by each mm-interferometer (\eg{} spatial
frequencies between 0 and 15\m{} for \PdBI{}). Those pseudo-visibilities
were merged with the observed, interferometric ones.  For imaging, we
followed the method described by Pety et al. (2005) to produce images in
different lines of the same source.  This results in inhomogeneous noise.
In particular, the noise strongly increases near the edges of the field of
view. To limit this effect, the mosaic field--of--view is truncated.

Moreover, the natural synthesized beam ($3.58''\times 1.89''$ at PA
37$\degr$) is very asymmetric due to the low declination of the Horsehead
nebula. We chooe to taper the weights of the $uv$ visibilities before
imaging using a Gaussian of size $400\m \times 115\m$ at PA 80$\degr$ to
obtain an almost circular clean beam ($3.65'' \times 3.34''$ at PA
48$\degr$). This considerably eased the deconvolution as 1) the larger beam
increases the brightness sensitivity and 2) the secondary sidelobes of the
dirty beam are much less patchy. Deconvolved image nevertheless still shows
low--level, fake, patchy structures at the scale of the clean beam, mainly
along the vertical direction. This is a known artifact of the H\"ogbom CLEAN
algorithm when a large spatial dynamic (field--of--view/resolution $\sim
110/3.5 = 30$) is combined with high enough signal--to--noise ratio.

\FigMeanCuts{} 

\FigSpectraCuts{} 

%%%%%%%%%%%%%%%%%%%%%%%%%%%%%%%%%%%%%%%%%%%%%%%%%%%%%%%%
\begin{table}
\caption{IRAM--30m line observation parameters from Gaussian fits.}
\begin{center}
\begin{tabular}{l c c c} 
\hline \hline  
Molecule/         & Offset   & $\Delta$v$_{FWHM}$ & $\int$T$_{A}^{*}$ $dv$  \\
Transition	  & ($''$)   & (km~s$^{-1}$) &  (K~km~s$^{-1}$)     \\ \hline \hline
CS $J$=2--1      & -52,-40  & 0.75$\pm$0.01  & 2.60$\pm$0.01\\
                  & -64,+30  & 0.89$\pm$0.01  & 3.63$\pm$0.01\\
	          & -35,-25  & 0.78$\pm$0.01  & 3.68$\pm$0.01\\
                  & -20,-15  & 0.77$\pm$0.01  & 3.55$\pm$0.01\\ \hline
CS $J$=3--2       & -52,-40  & 0.72$\pm$0.01  & 1.82$\pm$0.02\\
                  & -64,+30  & 0.93$\pm$0.01  & 2.58$\pm$0.02\\
	          & -35,-25  & 0.76$\pm$0.01  & 2.73$\pm$0.02\\
                  & -20,-15  & 0.80$\pm$0.01  & 2.40$\pm$0.02\\ \hline
CS  $J$=5--4      & -52,-40  & 0.43$\pm$0.02  & 0.35$\pm$0.01\\
                  & -64,+30  & 0.76$\pm$0.02  & 0.62$\pm$0.02\\
	          & -35,-25  & 0.58$\pm$0.02  & 0.52$\pm$0.01\\
                  & -20,-15  & 0.60$\pm$0.03  & 0.40$\pm$0.02\\ \hline
C$^{34}$S $J$=2--1& -52,-40  & 0.47$\pm$0.02  & 0.26$\pm$0.01\\
                  & -64,+30  & 0.67$\pm$0.04  & 0.38$\pm$0.02\\
	          & -35,-25  & 0.59$\pm$0.02  & 0.40$\pm$0.01\\
	          & -20,-15  & 0.58$\pm$0.03  & 0.45$\pm$0.02\\ \hline
C$^{34}$S $J$=3--2& -52,-40  & 0.48$\pm$0.04  & 0.20$\pm$0.01\\
                  & -64,+30  & 0.74$\pm$0.05  & 0.28$\pm$0.02\\ \hline
HCS$^+$ $J$=2--1  & -52,-40  & 0.8$\pm$0.3  & 0.07$\pm$0.03\\
                  & -35,-25  & 0.6$\pm$0.3  & 0.05$\pm$0.03\\
	          & -20,-15  & 0.9$\pm$0.4  & 0.10$\pm$0.03\\ \hline
\end{tabular}
\label{tab-summary_obs}
\end{center} 
\end{table} 

%%%%%%%%%%%%%%%%%%%%%%%%%%%%%%%%%%%%%%%%%%%%%%%%%%

\section{Results}

The PdBI integrated emission map of the CS $J$=2--1 line,
complemented with previous maps of CO $J$=1--0, 2--1 and C$^{18}$O $J$=2--1 lines
(Pety et al. 2005a), 
is shown in Fig.~\ref{fig:maps1}. Integrated emission profiles for these lines
and for the small hydrocarbon $c$--C$_3$H$_2$ 2$_{12}$--1$_{01}$ line emission and 1.2~mm 
continuum emission (Pety et al. 2005a) are shown in  Fig.~\ref{fig:cuts:mean}.
 In Fig.~\ref{fig:maps1} the four panels  are shown in  a coordinate
system adapted to the source: \ie{} maps have been rotated by 14\deg{} counter--clockwise 
around the image center to bring the exciting star direction in the horizontal direction as
 this eases the comparison with edge--on PDR models. Maps have also been horizontally shifted
 by $20''$ to the east in order to set the horizontal zero at the PDR edge
(delineated by the vertical line). Therefore, the coordinate system is converted from 
$\delta RA$ to $\delta x$  and from  $\delta DEC$ to $\delta y$ in arcsec 
(see Fig.~\ref{fig:maps1}). In the following, our spatial scale to interpret the PdBI maps 
will refer to the $\delta x$ scale.

PdBI observations show that the CS emission is strikingly different from that of other 
species observed at comparable spatial resolutions. CS  does not follow the filamentary
pattern along the cloud edge revealed by mid--IR (Abergel et al. 2003), H$_2$ 
(Habart et al. 2005) or CCH and c-C$_3$H$_2$ emission (Pety et al. 2005a). 
Instead, the behavior of the  CS $J$=2--1 line emission is more similar to that of CO 
and shows a steep rise across the PDR and a plateau in  the shielded region. Besides, 
the CS $J$=2--1 line emission peak occurs
in the well-shielded regions and does not coincide with the C$^{18}$O $J$=2--1 nor with 
the 1.2~mm continuum peaks (i.e. the temperature weighted density peaks, see 
Fig.~\ref{fig:cuts:mean}). Therefore, PdBI observations suggests
that CS is more abundant in the lower density regions and/or  it shows line opacity effects
in the denser regions. These results confirm that  there are differences  
in the spatial distribution of small hydrocarbons (Pety et al. 2005a) and other species
with similar excitation requirements.
PdBI CO, C$^{18}$O and CS line spectra along the direction of the exciting star 
($\delta y = 0''$) are shown in Fig.~\ref{fig:cuts:spectra}. The CS $J$=2--1 emission 
peak and line widths are comparable to those of C$^{18}$O~$J$=2--1. Nevertheless, 
the CS emission distribution profile in the $\delta x$ direction is less smooth,
 which can be an 
artifact  of the data reduction 
and/or related to the  CS photochemistry. The CS emission increases like the
C$^{18}$O emission in the PDR edge but in the shielded gas, CS rises when C$^{18}$O
decreases. On the other hand, $^{12}$CO peaks closer to 
the PDR edge and lines have  broader line widths (Pety et al. 2005a) indicative of their 
much larger opacities.

CS and C$^{34}$S single--dish observations at larger spatial scales are presented in 
Fig.~\ref{fig:obs_picoveleta}. CS line ratios are similar in all observed PDR positions.
 However,  there is a trend for CS lines to peak where C$^{18}$O emission decreases.
Again, this is indicative of larger abundances in the lower density regions and/or
line opacity effects in the denser regions. The latter hypothesis is playing a role because
CS $J$=3--2 lines  show asymmetrical profiles in the whole region, especially  red--wing
like self--absorptions. See for example the  $(-52,-40)$ position  with respect to the 
IRAM--30m C$^{18}$O~$J$=2--1 map of  Hily--Blant et al. (2005; Fig.~\ref{fig:obs_picoveleta}). 
In addition, CS $J$=3--2 and 2--1 lines must be optically thick since their intensity is 
only a factor $\sim$5 stronger than the analogous C$^{34}$S lines, 
significantly lower than the $^{32}$S/$^{34}$S=23 solar isotopic ratio
(Bogey et al. 1981). % Chemical Physics Letters 1981, Volume 81, Issue 2, p.256-260
In addition, CS $\frac{3-2}{2-1}$ line ratios are $\sim$0.7 while
the optically thin C$^{34}$S $\frac{3-2}{2-1}$ line ratios are larger $\sim$0.9. Therefore, 
the CS $J$=3--2 line is likely to be the most opaque CS line. 
Finally, Fig.~\ref{fig:obs_hcsp} shows 
clear detections of the HCS$^+$ $J$=2--1 line. As the expected HCS$^+$ abundance  is lower 
than that of C$^{34}$S, these lines are weak and should be optically thin. 
Line intensities are quite  similar in all observed PDR positions. 

\FigObsPicoVeleta{}

%%%%%%%%%%%%%%%%%%%%%%%%%%%%%%%%%%%%%%%%%%%%%%%%%
\begin{table}
\caption{Einstein coefficients, transition upper level energies  and critical densities 
 for the range of temperatures considered in this work.}
\begin{center}
\begin{tabular}{l c c c c} 
\hline \hline  
Molecule   & Transition&   $A_{ij}$              &  $E_i$      &  $n_{cr}$ 	               \\
           &           &  (s$^{-1}$)             &  (K)        &  (cm$^{-3}$)         \\ \hline \hline
C$^{18}$O  & $J$=2--1  &  6.01$\times$10$^{-7}$  &   15.8      &$\sim$8$\times$10$^{3}$ \\
HCS$^+$    & $J$=2--1  &  1.11$\times$10$^{-5}$  &    6.1      &$\sim$5$\times$10$^{4}$   \\
C$^{34}$S  & $J$=2--1  &  1.60$\times$10$^{-5}$  &    6.9      &$\sim$4$\times$10$^{5}$   \\
           & $J$=3--2  &  5.79$\times$10$^{-5}$  &   13.9      &$\sim$1$\times$10$^{6}$   \\ 
CS	   & $J$=2--1  &  1.68$\times$10$^{-5}$  &     7.1     &$\sim$4$\times$10$^{5}$   \\
           & $J$=3--2  &  6.07$\times$10$^{-5}$  &   14.1      &$\sim$1$\times$10$^{6}$   \\
           & $J$=5--4  &  2.98$\times$10$^{-4}$  &    35.3     &$\sim$5$\times$10$^{6}$ \\ \hline
\end{tabular}
\label{tab-n_cr}
\end{center} 
\end{table} 
%%%%%%%%%%%%%%%%%%%%%%%%%%%%%%%%%%%%%%%%%%%%%%%%%

\section{Numerical methodology}

\subsection{Photochemical models}
\label{subsec-PDR-mods}

We have used the \textit{Meudon PDR code} (publicly available at
 \textbf{http://aristote.obspm.fr/MIS/}), a photochemical model of a unidimensional 
stationary PDR (Le Bourlot et al. 1993).  The model has  been described in detail
elsewhere (Le Petit et al. 2006). In few words,  
the PDR code solves the FUV radiative transfer in an absorbing 
and diffusing  medium of gas and dust. This allows the explicit computation of the 
FUV radiation field (continuum+lines) and therefore, the explicit integration of 
consistent  C and S photoionization  rates together with H$_2$, CO, $^{13}$CO, and 
C$^{18}$O photodissociation rates. Penetration of  FUV radiation into the cloud strongly
 depends on dust properties through dust absorption and scattering of FUV photons. 
Properties of dust grains  are those described in Pety et al. (2005). We have taken a 
single dust albedo coefficient of 0.42 and an scattering asymmetry parameter of 0.6.

Once the FUV field has been determined, the steady-state chemical abundances are computed
 for a given chemical network. The  \textit{Ohio State University (osu)} gas--phase 
chemical network (osu.2005; September 2005 release; 
\textbf{http://www.physics.ohio-state.edu/$\sim$eric/research.html})
has been used as our chemical framework. The most important 
changes compared to previous versions are the decrease, by a factor of 2, of rate 
coefficients of photoionization and photodissociation reactions produced by 
cosmic--ray--induced H$_2$ secondary photons, the inclusion of fluorine (F) and its  
chemistry (see Neufeld et al. 2005) and the update of several reaction rates. 
In addition, several changes have been carried out by us on the chemical network.  
In particular, we have introduced different $^{18}$O bearing species into the chemical network
by assuming similar reaction rates to those involving the major isotopologues.
Fractionation reactions have been added following Graedel et al. (1982) and specific
photodissociation cross--sections for C$^{18}$O are explicitly introduced to compute
the corresponding photodissociation rate. When available, we have also used the
photodissociation rates given by van Dishoeck (1988), which are explicitly calculated 
for the Draine interstellar radiation field (ISRF).  Finally, we have further upgraded 
the sulfur network by adding the most recent reaction rates, dissociation channels and 
branching ratios
of HCS$^+$ and OCS$^+$ dissociative recombination (Montaigne et al. 2005) 
and by including the CS photoionization (ionization potential $\sim$11~eV).
These processes have direct impact on CS chemistry. The resulting network involves 
$\sim$450 species and $\sim$5000 reactions. Finally, the model computes the thermal 
structure of the PDR by solving the  balance between the most important processes heating
and cooling the gas (see Le Bourlot et al. 1993).
Our \textit{standard conditions} for the model of the Horsehead PDR 
include a power--law  density profile (Eq. 2) and a FUV radiation field enhanced by
a factor $\chi=60$ with respect  to the Draine ISRF (see table~\ref{tab-pdr_std_par}).
Different sulfur gas phase abundances, S/H, have been investigated.
To be consistent with  PdBI CO observations,
thermal balance was solved until the gas temperature
reached a minimum value of 30~K, then a constant temperature was assumed.

%%%%%%%%%%%%%%%%%%%%%%%%%%%%%%%%%%%%%%%%%%%%%%%%%
\begin{table}
\caption{Horsehead standard conditions and gas phase abundances.}
\begin{center}
\begin{tabular}{l c } 
\hline \hline  
Parameter                                 & Value                                        \\ \hline \hline
Radiation field $\chi$                    & 60 (Draine units)                      \\
Cosmic   ray ionization rate $\zeta$      &  $5 \times 10^{-17}$s$^{-1}$   \\ 
Density profile $n_H = n(H) + 2n(H_2)$    &  50 to  $2 \times 10^5$ cm$^{-3}$   \\
Line of sight spatial depth $l_{depth}$   &  0.05--0.1 pc           \\
Line of sight inclination angle $\varphi$ &  0$^o$ to 5$^o$ \\ 
He/H=$n(He)/n_H$                          &  $1.00 \times 10^{-1}$                \\
O/H                                       &  $3.02 \times 10^{-4}$  \\
C/H                                       &  $1.38 \times 10^{-4}$  \\
N/H                                       &  $7.95 \times 10^{-5}$ \\
%$S/H= n(S)/n_H$                          &  $10^{-8}$ to $2 \times 10^{-5}$ \\
$^{18}$O/H                                &  $6.04 \times 10^{-7}$ \\
Cl/H                                      &  1.00 $\times$ 10$^{-7}$ \\
Si/H                                      &  $1.73 \times 10^{-8}$ \\
Mg/H                                      &  $1.00 \times 10^{-8}$ \\
F/H                                       &  $6.68 \times 10^{-9}$ \\
Na/H                                      &  $2.30 \times 10^{-9}$ \\
Fe/H                                      &  $1.70 \times 10^{-9} $ \\
P/H                                       &  $9.33 \times 10^{-10}$ \\ \hline
\end{tabular}
\label{tab-pdr_std_par}
\end{center} 
\end{table} 
%%%%%%%%%%%%%%%%%%%%%%%%%%%%%%%%%%%%%%%%%%%%%%%%%

\subsection{Radiative transfer models}

We have used a simple nonlocal non--LTE radiative transfer code to model
our millimeter line observations. The code handles spherical and 
plane-parallel geometries and accounts for line trapping, collisional 
excitation, and radiative excitation by absorption of microwave cosmic 
background and dust continuum photons. Arbitrary density, temperature
or abundance profiles, and complex velocity gradients can be included.
A more detailed description is given in the Appendix. 
The choice of a nonlocal treatment is needed to analyze 
optically thick lines of abundant, high--dipole moment molecules, such as CS, 
in regions where  the gas density is below the critical densities
of the associated transitions. 
Table~\ref{tab-n_cr} shows the critical densities of observed 
C$^{18}$O, HCS$^+$ and CS lines. 
Under these conditions, radiative transfer and opacity
effects  may dominate the line profile formation. 
%In the most extreme cases, not taking into account these effects 
%can lead to unrealistic abundances and/or physical conditions.
Our radiative transfer analysis has been used to infer abundances and
physical conditions directly from observations but also to predict
line spectra from the photochemical model results.
%The model has been used for CS, C$^{34}$S, HCS$^+$ and C$^{18}$O.
The following temperature dependent collisional rate 
coefficients\footnote{Some of them retrieved
from BASECOL, a data base for collisional excitation data
at \texttt{http://www.amdpo.obspm.fr/basecol}.
We considered  H$_2$, He and H as the collisional
partners in all CS, C$^{34}$S, HCS$^+$ and C$^{18}$O excitation  models.
 See Appendix.} 
have been adopted:\\\\  
-- for CS: we have used the latest CS + He collisional rates from 
Lique  et al. (2006), kindly provided by F.~Lique prior to publication,
scaled by the reduced mass factor $(\mu_{CS-H_2}/\mu_{CS-He})^{1/2}$. 
Most of the models were repeated with the older collisional rates
of Turner et al. (1992).\\ 
-- for C$^{34}$S: same as CS but using C$^{34}$S spectroscopy to compute
collisional excitation rates through detailed balance.\\ 
-- for C$^{18}$O: CO + H$_2$  de--excitation 
rates from Flower (2001) but using  C$^{18}$O spectroscopy to compute
collisional excitation rates through detailed balance.\\
-- for HCS$^+$: HCS$^+$ + He collisional rates from Monteiro (1984),
scaled by the reduced mass factor $(\mu_{HCS^+-H_2}/\mu_{HCS^+-He})^{1/2}$,
have been used. 
%We have also checked with
%HCO$^+$ + H$_2$ de--excitation  rates  (Flower et al. 1999) corrected
%by the reduced mass  factor $(\mu_{HCO^+-H_2}/\mu_{HCS^+-H_2})^{1/2}$
%and  HCS$^+$ spectroscopy to compute
%collisional excitation rates through detailed balance. 

\section{Modeling and interpretation}

\subsection{CS, C$^{34}$S and HCS$^+$ single--dish emission}
\label{CS_exc}

In order to get a first order approximation of the  CS excitation and column 
density, we have assumed that level populations are only determined 
by a Boltzmann distribution at a single rotational temperature.
If one accepts that lines are optically thin, this approach 
corresponds to the widely used rotational--diagram.
However, observed CS/C$^{34}$S intensity ratios, and CS line profiles
(see Fig.~\ref{fig:obs_picoveleta}) clearly show that the  low--$J$ CS lines 
are optically thick towards the Horsehead.
Therefore, we have included optical depth effects 
and built a rotational--diagram corrected for opacity
through:
\begin{equation}
ln\,\frac{N_{i}^{thin}}{g_i} + ln\,C_{\tau}
= ln\,N-ln\,Q- \frac{E_i}{T_{rot}}
\label{eq-tot-diag}
\end{equation}   
where $N_{i}^{thin}$ are the upper level populations determined
from observations in the optically thin limit (underestimated
if lines are optically thick), $E_i$ is the upper $i$-level energy in K,
$Q$ is the partition function at T$_{rot}$ and $C_{\tau}$ is  the line opacity correction
 factor 
$\frac{\tau_{ij}}{1-e^{-\tau_{ij}}}>$1 (Goldsmith \& Langer 1999). We have performed this
 analysis at different cloud positions. Resulting diagrams are shown in 
Fig.~\ref{fig:ROTdiag} as a function of different CS $J$=2--1 line opacities 
($\tau_{2-1}=0,1$ and 5). In the optically thin limit CS column densities are 
$N(CS)$ $\sim$5$\times$10$^{13}$~cm$^{-2}$ and they have  to be considered as 
 lower limits. %However, even assuming that the CS excitation is correctly represented 
%by a Boltzmann
% distribution,  the $N(CS)$ uncertainty can be a factor $\sim$10 depending on 
%line optical depths (this factor can be higher for molecular
%clouds where the column density of material is larger than in PDRs).
\FigRotdiag{}
Low excitation temperatures (T$_{rot}\sim$6-9~K) are also inferred from the 
rotational--diagrams. These values, much lower than expected gas temperatures 
in a PDR, are suggestive of radiative excitation effects in CS lines and level 
populations far from thermalization. Therefore, we only use the rotational--diagrams
 as input for the first iteration of a  more complex analysis.

In order to obtain a more detailed overview of the CS excitation, we have made 
several statistical equilibrium calculations (see Appendix) around the expected
physical conditions in the Horsehead. In particular, we have run a grid
of single--component models for T$_k$=10, 20, 30, 50 and 70~K,
$n(H_2)$=10$^4$, 5$\times$10$^4$, 10$^5$ and 5$\times$10$^5$~cm$^{-3}$, and
$\chi(CS)$ from 10$^{-10}$ to 10$^{-7}$. As a reference value, the cloud 
total extinction is assumed to be constant and equal to $A_V$=20~mag in
all models, i.e. the spatial length is changed accordingly.
Fig.~\ref{fig:MTCgrid} specifically shows selected results for T$_k$=30~K,
which gives appropriate absolute intensities for the CS lines.
In particular, integrated line intensity ratios of observed lines
as a function of CS abundance for different densities are shown.
Averaged ratios from CS single--dish observations are  
$\frac{3-2}{2-1}$ $\sim$0.7, 
$\frac{5-4}{2-1}$$\sim$0.2 and $\frac{5-4}{3-2}$$\sim$0.3.
Therefore, densities $\geq$5$\times$10$^4$~cm$^{-3}$
are needed to populate the CS intermediate--$J$ levels.
%otherwise the CS $J$=5--4 line would stay undetected. 
On the other hand, for high densities ($\geq$5$\times$10$^5$~cm$^{-3}$), collisions start
to efficiently populate these levels and the predicted line ratios
involving the CS $J$=5--4 line become much larger than observed.
Thus, mean densities are in the range  $n(H_2)\simeq$(0.5--1.0)$\times$10$^5$~cm$^{-3}$, i.e.
lower than CS critical densities (table~\ref{tab-n_cr}).
Excitation temperatures are predicted to be subthermal, T$_{ex}$ $<$T$_k$, especially
for the highest frequency lines. Due to line--trapping, the maximum T$_{ex}$ is
reached at the center of the cloud, while it graduately drops at both  surfaces
where line photons are optically thin and line trapping is not efficient. 
The only exception is the CS $J$=1--0 transition which shows an increase
of the excitation temperature, T$_{ex}^{1-0}$, at both  surfaces.
This rising in T$_{ex}^{1-0}$ is due to the significant collisional excitation
coupling from the $J$=0 to $J$=2 level, and to the large radiative  de--excitation rates
from $J$=2 to $J$=1 level. At both surfaces, where optically thin CS $J$=2--1
line photons can easily escape from the cloud (T$_{ex}^{2-1}$ decreases),
the above excitation conditions favor the population of the $J$=1 
level with respect to the $J$=0 level. Therefore, T$_{ex}^{1-0}$ can reach 
large suprathermal values. A typical example is shown in Fig.~\ref{fig:CS_bench}.
Thus, within this range of parameters and even if physical conditions are homogeneous, 
excitation gradients must be taken into account.

\FigMTCgridCS{}

For these temperatures and densities, T$_k$ $\simeq$20--30~K and 
$n(H_2)$ $\simeq$(0.5--1.0)$\times$10$^5$~cm$^{-3}$, the CS $\frac{5-4}{2-1}$ 
and $\frac{5-4}{3-2}$ line ratios are better fitted in the interval
 $\chi(CS)$ $\simeq$(0.2-1.0)$\times$10$^{-8}$. Nevertheless, the $\frac{3-2}{2-1}$ 
line ratio is systematically predicted to be larger than observed in these single--component
 models.
Therefore, a more complex density structure and/or additional opacity effects in 
 low--$J$ CS lines may be affecting the observed profiles.
The latter hypothesis is clearly favored by the fact that the C$^{34}$S $\frac{3-2}{2-1}$ line 
intensity ratio is larger ($\sim$0.9) than the CS $\frac{3-2}{2-1}$ ratio ($\sim$0.7) and 
thus closer to the single--component  model predictions. Since the C$^{34}$S emission 
is optically thin, radiative transfer effects are less important and  C$^{34}$S column
densities can be  accurately determined.

C$^{34}$S single--component radiative transfer models for selected positions within the 
region have been run (Fig.~\ref{fig:mods_cs}). Since C$^{34}$S emission can
arise from different gas components of higher density, not resolved 
by the large single--dish beam, we have modeled each position in spherical geometry. 
This allows to explore different components of different beam filling factors.
The maximum extinction in the models varies from A$_V$=20 to 12~mag depending on the 
cloud position. These values are consistent with those obtained from single--dish 1.2~mm 
dust continuum emission observations (Teyssier et al., 2004, Pety et al. 2005a).
Following our previous general excitation calculations we have considered gas temperatures
 in the range 20--25~K. For these conditions, densities between $n(H_2)$=7$\times$10$^4$ 
and  1.2$\times$10$^5$~cm$^{-3}$ satisfactorily reproduce the observed C$^{34}$S absolute 
intensities.  Best fits are obtained for turbulence velocities 
(see Appendix for the definition of v$_{turb}$)
between 0.3 and 0.4~km~s$^{-1}$ (Table~\ref{tab-RT_single}). Although C$^{34}$S is
slightly enhanced where C$^{18}$O 
decreases, we have averaged the 4  positions to find the mean C$^{34}$S abundance in the 
region covered with  single--dish observations and found 
$\chi(C^{34}S)$=(3$\pm$1)$\times$10$^{-10}$.
Since nucleosynthesis models favor a constant galactic $^{32}$S/$^{34}$S ratio 
and  many observations reproduce the solar ratio within their error bars (Wannier et al. 1980 ; 
Frerking et al. 1980), especially in local diffuse clouds (Lucas \& Liszt, 1998),  
we adopt $^{32}$S/$^{34}$S=23 here as the isotopic ratio in the Horsehead. 
Therefore, the derived $\chi(C^{34}S)$ abundance  translates to
 $\chi(CS)$=(7$\pm$3)$\times$10$^{-9}$.
The same physical conditions at each position have been used to model
the HCS$^+$ $J$=2--1 lines (see Fig.~\ref{fig:obs_hcsp}). Lines
are reproduced for an averaged abundance of 
$\chi(HCS^+)$=(4$\pm$2)$\times$10$^{-11}$, therefore, 
a CS/HCS$^+$ $\simeq$175 abundance ratio is derived. 
%Note that larger turbulence velocities, $\geq$0.3~km~s$^{-1}$, are required
%to correctly fit the HCS$^+$ lines. If this is not a consequence of the
%lower quality of HCS$^+$ observations (see error bars in 
%table~\ref{tab-summary_obs}), different spatial
%origins for the dominant C$^{34}$S and HCS$^+$ may contribute to the observed
%line widths differences. It is 
%likely that HCS$^+$ traces the surface, and more turbulent  PDR layers,
%while C$^{34}$S can dominate in the denser clumps interiors.
%In the next section we analyze these results in the context of 
%sulfur chemistry.

%%%%%%%%%%%%%%%%%%%%%%%%%%%%%%%%%%%%%%%%%%%%%%%%%%%%%%%%
\begin{table}
\caption{One-component radiative transfer model parameters.}
\begin{center}
\begin{tabular}{l c } 
\hline \hline  
Parameter	   &  Value  \\ \hline \hline
T$_{k}$	           &  20--25 K\\
$n(H_2)$           &  (7--12)$\times$10$^{4}$ cm$^{-3}$\\
v$_{turb}$         &  0.3--0.5 km s$^{-1}$\\
$\chi(C^{34}S)$    &  (3$\pm$1)$\times$10$^{-10}$\\
$\chi(HCS^{+})$    &  (4$\pm$2)$\times$10$^{-11}$\\  \hline
\end{tabular}
\label{tab-RT_single}
\end{center} 
\end{table} 
%%%%%%%%%%%%%%%%%%%%%%%%%%%%%%%%%%%%%%%%%%%%%%%%%%%%%%%%

\FigModsCS{}

\FigObsHCSP{}

Using the CS abundance inferred from the C$^{34}$S analysis, we have now tried to fit
 the CS lines at each position. Since a single--component model does not reproduce 
 the observed line ratios and absolute  intensities, we have explored other possibilities. 
In principle, CS low--$J$ lines are  optically thick and may not trace the high density gas 
revealed  by C$^{34}$S, especially if the medium is inhomogeneous and dense clumps and a
more diffuse interclump medium coexist. The same argumentation has been used to interpret 
HCN and  H$^{13}$CN observations in the Orion Bar PDR (Lis \& Schilke, 2003).
In addition, it is well known that low--$J$ CS lines may not be a good column density tracer 
if their emission is scattered by a low density halo (Gonz\'alez-Alfonso \& Cernicharo 1993). 
This process can be a common effect in optically thick lines of high--dipole moment molecules 
 such as CS or  HCO$^+$ (Cernicharo \& Gu\'elin 1987). 
In this scenario, the CS $J$=3--2 and 2--1 lines from the dense medium will
be attenuated and scattered over larger areas than the true spatial extend of
the dense clumps. This possibility has been analyzed in more detail in the next section.
Fortunately, observations of the CS $J$=5--4 line allow to directly trace the dense clumps 
more safely  (Table~\ref{tab-RT_CS}). In particular, we found that these lines can only be 
reproduced with denser gas components,  $n(H_2)$= (4$\pm$2)$\times$10$^5$~cm$^{-3}$,
not resolved by the $\sim$10$''$ beam of the IRAM--30m telescope at $\sim$250~GHz.
Note that the CS $J$=5--4 line widths are fitted if the turbulent velocity in the denser
 gas is $\sim$0.2~km~s$^{-1}$, a factor 2 lower than the one required by the
C$^{34}$S $J$=3--2 and 2--1 lines (Fig.~\ref{fig:mods_cs}).
Thus, a different spatial origin for this line emission is favored.

At this stage  we have a general knowledge of the CS and HCS$^+$ excitation and abundance in
 the region. In the following sections we concentrate in the photochemistry of these species.
 Only higher angular observations provide the appropriate linear scale to resolve the most
 important physical gradients in the PDR edge.  Hence, interferometric observations  should
 allow  a better comparison with chemical predictions.

%%%%%%%%%%%%%%%%%%%%%%%%%%%%%%%%%%%%%%%%%%%%%%
\begin{table}
\caption{Two--component radiative transfer model parameters.}
\begin{center}
\begin{tabular}{l c } 
\hline \hline  
Parameter	    &  Value  \\ \hline \hline
T$_{k}$	            &  20--25 K\\
$n(H_2)$            &  (3--7)$\times$10$^{4}$ cm$^{-3}$\\
\hspace{0.2cm} dense component    &  (2--6)$\times$10$^{5}$ cm$^{-3}$\\
\hspace{0.2cm} (filling factor)   &  0.3        \\
v$_{turb}$          &  0.3--0.4 km s$^{-1}$\\
\hspace{0.2cm} dense component    &  0.2--0.3 km s$^{-1}$\\
$\chi(CS)$          &  (7$\pm$3)$\times$10$^{-9}$\\
S/H               &  (3.5$\pm$1.5)$\times$10$^{-6}$\\  \hline
\end{tabular}
\label{tab-RT_CS}
\end{center} 
\end{table} 
%%%%%%%%%%%%%%%%%%%%%%%%%%%%%%%%%%%%%%%%%%%%%%%%%%%%%%%%

\subsection{The PDR edge} 

PdBI C$^{12}$O $J$=2--1, 1--0, C$^{18}$O $J$=2--1, and CS $J$=2--1 observations along the
 direction of the exciting star (at $\delta y$= 0$''$) are shown in 
Fig.~\ref{fig:cuts:spectra}. Here we take these spectra as representative of the PDR edge 
and try to constrain its physical conditions through a combined analysis of photochemical 
and radiative transfer models. Both models use a unidimensional plane--parallel description 
of the geometry. Although some physical processes 
require more complex geometries, the main physical and chemical gradients 
across the illuminated direction can be consistently described in this way. 
Plane--parallel geometry was judged to be the best approach for
this edge--on PDR since  H$_2$ and PAH emissions are only observed at the illuminated edge
and not deeper inside the cloud (Habart et al. 2005).

In this analysis, we have used the PdBI  CS $J$=2--1 and C$^{18}$O $J$=2--1 lines.
As low--$J$ $^{12}$CO optical depths are very high, they do not trace the bulk of material.
The intensity peak of these lines only provide a good estimation of the CO excitation 
temperatures (i.e. a lower limit to the gas temperature). Since the asymptotic brightness
 temperature of CO $J$=1--0 lines is $\sim$30~K, we take this value as the minimum
of T$_k$ in the PDR. We note that lower temperatures do not reproduce
the observed  line intensities. For the rest of the (warmer) positions closer to the PDR edge,
 the gas temperature  was determined by solving the thermal balance.
The predicted gas temperature in the density peak is $\sim$50~K while it rises to
 $\sim$200--250~K in the H$_2$ emitting regions where the density is
 $n_H$ $\simeq$10$^3$-10$^4$~cm$^{-3}$. More exact  temperature values 
require observations of higher--$J$ CO lines at comparable spatial resolution. 
We are currently analysing $^{13}$CO $J$=3--2 data from the SMA interferometer.

%\clearpage

Regarding the density structure, both the observed H$_2$ and PAH mid--IR emission, together
 with their spatial segregation, are much better reproduced with a steep density gradient
 than with an uniform density (Habart et al. 2005). 
The same  density gradient is needed to correctly reproduce the observed offset between the 
small hydrocarbons (Pety et al. 2005a) and H$_2$ emission (where the density
is not at its peak). 
Therefore, in order to reproduce PdBI observations of CS and C$^{18}$O, a steep 
power--law density gradient at the illuminated regions and a step--density in the more
 shielded region have been assumed. 

The following methodology was carried out: a full PDR model with
Horsehead \textit{standard conditions} (see section~\ref{subsec-PDR-mods})
was run with a particular choice of the density gradient described
in Eq.~\ref{eq-density_grad}. 
Afterwards, the PDR output was used as input
for the nonlocal radiative transfer calculation in a fashion described
in appendix \ref{sub-apendix2}. 
In this way, physical parameters can be tuned more accurately
by iteration of different radiative transfer models. 
Once better parameters have been found, a new PDR computation is 
performed with this choice of physical parameters.
Hence, the most appropriate physical and chemical
description of the PDR edge was found through the
\textit{PDR model$\rightarrow$transfer model$\rightarrow$check with 
observations$\rightarrow$transfer model$\rightarrow$PDR model}
iterative process.  Therefore, synthetic CS and C$^{18}$O abundance
profiles are consistently computed as a function of the edge distance
$\delta x$ (in arcsec) and directly compared with observations. 

\FigModsPDR{}

Different PDR spatial depths were investigated. Depending on the adopted density profile, 
the spatial depth $l_{depth}$ is determined by the line of sight visual extinction. However,
the A$_V$ value depends on the method used to measure it. If optically thin 1.2~mm
dust emission is used (Teyssier et al. 2004; Pety et al. 2005a; Habart et al. 2005), 
the resulting column densities depend on the usually unknown  grain properties and
on the assumed temperature.
 Taking into account our poor knowledge of the cloud thermal structure, a factor $\sim$2 of 
uncertainty in A$_V$ can be assumed.  In addition, the angular resolution of millimeter 
continuum observations is at least a factor $\sim$2 worse than PdBI molecular 
line observations.  Due to the steep decrease of the density towards the edge, and due 
to the $\sim$11$''$ beam of 1.2~mm  continuum observations, the 
observed emission peak will appear deeper inside the cloud, shifted a few arcsec from 
the real density peak (which is closer to the edge). Therefore, together with the PDR edge 
location, the exact peak density position can also be uncertain by a few arcsec.
Finally, beam dilution has to be also taken into account when comparing single--dish 
versus interferometric observations. 
%In particular, the effective extinction in the 
%small--scale structure of the cloud will be clearly enhanced in PdBI observations. 
Here we have chosen $l_{depth}$ $=$0.05-0.1~pc, which implies  extinction peaks around
 A$_V$ $\simeq$15-30~magnitudes. These values are expected in compact globules
(Reipurth \& Bouchet 1984). Since CS and C$^{18}$O excitation and line transfer
are quite different, the following combined analysis provides an accurate
description of the edge density structure.
The empirical density profile in the models, $n_H $=$n(H)$+2$n(H_2)$, as a function
of  $\delta x$ is:
\begin{equation}
n_H(\delta x)= \left\{
\begin{array}{c}
n_H(0)+[n_H(\delta x_1)-n_H(0)]\,\left(\frac{\delta x}{\delta x_1}\right)^\beta  
                                ;  \delta x_1 \geq \delta x \geq 0\\
n_H(\delta x_1)      ; \delta x_2 \geq  \delta x >  \delta x_1\\
\hspace{-0.9cm}n_H(\delta x_2)      ;   \delta x >  \delta x_2         \\
\end{array}
\right.
\label{eq-density_grad}
\end{equation}   
where $\delta x$ is the distance away from the PDR edge, $n_H(0)$ is the ambient density 
at the edge, and $n_H(\delta x_1)$ and $n_H(\delta x_2)$ are constant  densities in the
$\delta x_2 \geq  \delta x > \delta x_1$ and $ \delta x > \delta x_2$ regions respectively. 
Selected photochemical models are shown in Fig.~\ref{fig:ModsPDR}.
The normalized population of the H$_2$ v=1, $J$=3 level  is shown in the $upper$ panel
 and is used to place the $\delta x$--axis origin of the models and thus to accurately 
check with observations.  Although some uncertainty in the location the PDR edge exists, 
we place
the peak of this curve at the maximum of  observed H$_{2}$ 1--0 S(1) 2.12~$\mu$m  excited 
line  ($\delta x$ $\sim$10$''$; Habart et al. 2005). Best models are found for a peak density
around  $n_H(\delta x_1)$=2$\times$10$^5$~cm$^{-3}$. This density is reached in
a length of $\sim$2.5$''$--5$''$ (or 5--10$\times$10$^{-3}$~pc) and stays constant in a length 
of $\delta x_2 -\delta x_1$ $\lesssim$20$''$ (or 0.04~pc). 
In order to fit the smooth decrease of C$^{18}$O emission and also of the 1.2~mm 
continuum emission, the density has to decrease again by at least a factor $\sim$2. We have
 simply modeled this as a step-function for $\delta x$$>$$\delta x_2$ and decrease
the density to $n_H(\delta x_2)$=10$^5$~cm$^{-3}$.
We have chosen $\delta x_1 = 12''$ and $\delta x_2 =30''$.
Our models confirm that high density gas and a large  gradient slope, $\beta$ $\sim$3--4,
are needed to reproduce the PdBI and H$_2$ observations (Habart et al. 2005), although
we found a slightly smaller gradient scale length. 
%In the next section we discuss the possible origin 
%of such a steep slope.

As proposed by Habart et al. (2005) the PDR edge can be slightly inclined with respect to 
the line of sight by a small  angle $\varphi$. In plane--parallel geometry,  the maximum 
inclination can be estimated assuming that the observed spatial extend of the H$_2$ emission, 
$d_{H_2}$,  is mainly due to the projection of $l_{depth}$ in the plane of the sky, 
thus $sin\,\varphi\simeq d_{H_2}/l_{depth}$. Since $d_{H_2}$ $\sim$0.01~pc,
an inclination angle $\varphi$ $\sim$5$^o$, has been considered in the radiative transfer 
models (see Appendix \ref{sub-apendix2}). As expected, even such a small inclination shifts 
the emission peak significantly and should therefore be taken into account.
Fig.~\ref{fig:PdbImtc} shows the PdBI C$^{18}$O line observations and the combined 
PDR+transfer modeling including such geometrical effects. The agreement is excellent, 
probably favored by the well--established CO  photochemistry (Fig.~\ref{fig:ModsPDR})
and because C$^{18}$O $J$=2--1 lines do not show complex radiative transfer effects 
($\tau_{2-1}$ $\sim$0.8).

To analyse the spatial distribution of the CS abundance predicted by photochemical
models at the PDR edge we have also tried to fit the PdBI CS $J$=2--1 lines.  
Fig.~\ref{fig:ModsPDR} shows the effects of different sulfur abundances;
S/H=2$\times$10$^{-5}$ and  S/H=2$\times$10$^{-6}$. 
%We found that these kind of models
%(single physical conditions in a given line--of--sight)
%overestimate CS $J$=2--1 line intensities by a  factor of $\sim$4 for a solar sulfur
%abundance. Hence, the sulfur abundance should be around $\sim$10$^{-6}$. 
Fig.~\ref{fig:CSPdbImtca} (no inclination) and  Fig.~\ref{fig:CSPdbImtcb} (inclination considered)
show the resulting synthetic CS map, using S/H=2$\times$10$^{-6}$ %(see next section) 
and a minimum gas temperature of 30~K,
over  PdBI observations at two constant $\delta y$ cuts ($\delta y$=30$''$ and 0$''$).
Contrary to  C$^{18}$O, the CS emission detected with the PdBI at a fixed $\delta x$ near 
the edge shows an emission gradient in the $\delta y$ direction, 
e.g. line peaks are brighter as $\delta y$ increases. As a consequence
model predictions fit better the  $\delta y$=30$''$ cut than the $\delta y$=0$''$ one.
Besides, larger gas phase sulfur abundances 
are obtained if the bulk of the gas in the PDR edge is warmer,  i.e. minimum gas 
temperatures of $\sim$100~K (Fig.~\ref{fig:ModsPDR}, right panel). 
This may be an indication of larger temperatures at the cloud edge and lower
sulfur depletions. Note that an accurate estimation of the CS abundance at high
resolution  from a single PdBI  line is not straightforward.
Such determination requires aperture synthesis observations of 
additional CS lines to have a minimum idea of the CS excitation in
different positions.

\FigPdBIMTC{}
\FigCSPdBIMTCa{}
\FigCSPdBIMTCb{}
\FigHalos{}

In addition, since C$^{34}$S observations at the same high--angular  resolution were not 
available, we could not estimate additional opacity  effects in  previous PdBI
 CS models  ($\tau_{2-1}\gtrsim$2). 
In the following we have tried to estimate the worse possible scenario affecting the CS 
lines in the line--of--sight, i.e. the presence of a surrounding low density halo. Of course, 
the greatest effect can appear in the shielded regions where the gas column density is largest.
Therefore we modelled a typical position where CS is well spatially resolved 
with the following 
parameters: $l_{depth}$=0.1 pc,  T$_k$=30~K, $n$(H$_2$)=10$^5$~cm$^{-3}$,
 v$_{turb}$=0.35~km~s$^{-1}$  and $\chi$(CS)=7$\times$10$^{-9}$,
(the averaged CS abundance obtained from the detailed CS and C$^{34}$S
excitation  analysis of previous section). We consider in addition that a low density
halo of diffuse gas with the same $\chi$(CS) surrounds the region. 
We take T$_k$=10~K, v$_{turb}$=0.7~km~s$^{-1}$ and densities in the
interval $n$(H$_2$)=(5--10)$\times$10$^3$~cm$^{-3}$. The same modeling was carried
out for C$^{34}$S. Fig.~\ref{fig:Halos} shows model results. As expected, a low density halo
efficiently self-absorbs CS line photons in the most opaque lines, i.e. the low--$J$ CS lines. 
As a result, the observed CS line intensities are attenuated and abundances can be easily 
underestimated. However, this effect can be different at different positions, 
since the line opacity also changes.  Apart from uncertainties in sulfur chemistry or 
instrumental effects in interferometric observations, diffuse gas can also contribute
to explain differences between models and observations in 
Figs.~\ref{fig:CSPdbImtca} and  \ref{fig:CSPdbImtcb}.
Since, optically thick lines are affected by this effect (Fig.\ref{fig:Halos}), only 
the observation of $^{13}$CS or C$^{34}$S isotopologues can help to provide more accurate 
abundance determinations. 
In the following, the CS chemistry is analyzed in more  detail.

\subsection{CS chemistry and S-abundance} 

Predicted C$^{18}$O/CO/C/C$^+$ and CS/HCS$^+$/S/S$^+$ structures for 
a unidimensional PDR with Horsehead \textit{standard conditions} are 
shown in Fig.~\ref{fig:ModsPDR} (see section~\ref{subsec-PDR-mods}). 
Variation of the sulfur elemental abundance almost does not affect 
the CO or C$^{18}$O abundance profiles, but it slightly  modifies
the predicted C/C$^+$ abundance profiles because charge transfer
reactions between C$^+$ and S, and between C and S$^+$ are clearly altered by 
the sulfur depletion. 
The following results are of course determined
by our present knowledge and uncertainties on S--chemistry, and on
reaction rates at different temperatures.
According to the latest \textit{ion storage ring} experiments
(Montaigne et al. 2005), only
19\% of the HCS$^+$ dissociative recombination results in CS + H
while the fracture of the C--S bond dominates the dissociation (81\%).
Since these experiments can not separate the contribution of the
CH + S or SH + C channels in the latter process, we have adopted
the same branching ratio (0.405) for both channels.
The reaction rate coefficient 
is $k_{DR}(HCS^+)_{fast}$=9.7$\times$10$^{-7}$(T/300)$^{-0.57}$~cm$^3$~s$^{-1}$.
We have also included the latest OCS$^+$ dissociative recombination rates
from Montaigne et al. (2005). The CS + O production channel now occurs at
a rate 3 times slower than in previous chemical networks.
All these modifications clearly influence the amount of CS formed
from a given sulfur abundance, and thus the sulfur depletion 
estimations. Fig.~\ref{fig:ModsPDR} (left panel) also shows the effect of adopting the older
HCS$^+$ and OCS$^+$ dissociative recombination rates and branching ratios.
In particular, $k_{DR}(HCS^+)_{slow}$=5.8$\times$10$^{-8}$(T/300)$^{-0.75}$~cm$^3$~s$^{-1}$.
However, since $HCS^+ + e^- \rightarrow CS + H$ was the only channel considered and
the $OCS^+ + e^- \rightarrow CS + O$ process was  faster, smaller sulfur abundances
were required to obtain the same CS abundances.

In the most external layers of the cloud, still dominated by the FUV
radiation field, CS is predominantly formed  by HCS$^+$ dissociative
recombination and principally destroyed by photodissociation and charge transfer 
with H$^+$. Once the gas is shielded, OCS$^+$ dissociative
recombination and reaction of C with SO also contributes to CS formation,
while its destruction  is now governed by ion--molecule reactions, mainly with HCO$^+$ 
but also with H$_3$O$^+$. These last two reactions with abundant molecular ions
return HCS$^+$ again. The peak abundance of HCS$^+$ occur at A$_V$ $\lesssim$2~mag, 
where it is formed by reaction of CS$^+$ with H$_2$ and destroyed by dissociative recombination.  For this reason, an order of magnitude change in
$k_{DR}(HCS^+)$ clearly modifies its peak abundance in the outer PDR layers.
In the more shielded regions, HCS$^+$ destruction is dominated by dissociative
recombination and reaction with atomic oxygen to form
HCO$^+$ and OCS$^+$. Since the predicted CS abundance scales with S/H, and
CS formation is dominated by HCS$^+$ dissociative recombination,
we have used our CS/C$^{34}$S/HCS$^+$ observations and modeling
to estimate S/H.

Fig.~\ref{fig:cs_hcsp_pdr} shows results of a grid of photochemical
models for different sulfur elemental abundances from S/H=10$^{-8}$ to 2$\times$10$^{-5}$, 
using the latest HCS$^+$ and OCS$^+$ dissociative recombination rates.
CS and HCS$^+$ abundances with respect
to H$_2$ are shown as a function of S/H at two different PDR positions
(A$_v \sim$10 and $\sim$2~mag  respectively; see Fig.~\ref{fig:ModsPDR}). 
Densities at these positions are the same, $n(H_2)$=10$^{5}$~cm$^{-3}$,
but we  have taken different PDR positions in order to plot the 
HCS$^+$ maximum abundance and to get the CS/HCS$^+$ ratio closer to observations.
Inside the cloud, the predicted maximum HCS$^+$ abundances
are a factor $\sim$3 lower than observed. 
Horizontal shaded regions mark the CS and HCS$^+$ abundances 
derived from observations and radiative transfer modeling.
For clarity, HCS$^+$ abundances have been multiplied by a 
factor of 1000. Finally, the vertical shaded region shows the estimated sulfur 
elemental abundance in the  Horsehead derived from the overlap region 
between observed and predicted abundances. 
We derive S/H $\sim$(3.5$\pm$1.5)$\times$10$^{-6}$ as the mean value for the PDR.
Note that CS is used for the upper limit and HCS$^+$ for the lower limit.
However, according to the inferred HCS$^+$ abundance,
larger sulfur abundances are still  possible.
%and/or HCS$^+$ is underproduced in models compared to observations.

\FigPdrCSHCSP

\section{Discussion} 

Our multi--transition single--dish and  aperture synthesis observations and
modeling of CS and related species allow us to constrain the sulfur gas phase chemistry
in the Horsehead PDR and it  also gives some insights on the dense gas properties.

\subsection{Densities}
The densities found in this work, $n(H_2)\simeq$10$^5$~cm$^{-3}$, 
are larger to those inferred from previous studies based on single--dish CO observations 
(Abergel et al. 2003; Teyssier et al. 2004).
This may be the indication of an inhomogeneous medium characterized by a 
interclump medium (well traced by CO) and a denser clump medium
(better traced by high dipole molecules). Both high densities and inhomogeneous medium are common in other PDRs such as the Orion Bar (Lis \& Schilke 2003).
In particular, we have shown that  unresolved gas components up 
to $n(H_2)$ $\simeq$(2--6)$\times$10$^5$~cm$^{-3}$ are required to explain the CS $J$=5--4 line
 emission in the Horsehead. However, 
Abergel et al. (2003) did not find inhomogeneities in analysing ISOCAM images of
the Horsehead. Nevertheless, they noted that clumpiness at scales smaller than the
upper limit of the FUV penetration depth ($\sim$0.01~pc) could not be excluded.
Our best  models of the  CS $J$=5--4 line emission require an unresolved component
with a radius of  $\sim$5$\times$10$^{-3}$~pc. This component can of course be
further fragmented itself.
Nevertheless, it is difficult to distinguish between clumpiness at scales below 
$\sim$0.01~pc and the presence of a lower density envelope surrounding the cloud.
Since CO $J$=1--0 and 2--1 line opacities  easily reach large values, their observed  
profiles are formed in the very outer layers of the cloud and thus they can arise from
the most diffuse  gas ($n(H_2)$ $\sim$5$\times$10$^3$~cm$^{-3}$).
Interferometric observations of intermediate--$J$ lines of high dipole species
such as CS or HCO$^+$ will  help to clarify the scenario.

The high angular resolution provided by PdBI CS and C$^{18}$O observations
reveals that the Horsehead PDR edge is characterized by steep density gradient
rising from ambient densities to $n(H_2)$ $\sim$10$^5$~cm$^{-3}$ in
a length of $\sim$0.01~pc and kept roughly constant up to $\sim$0.05~pc, 
where the density decreases again at least a by factor 2. 
%As probed by Habart et al. (2005), 
%this density profile is needed to reproduce the H$_2$ fluorescent emission (where the density 
%is not at its peak) but also to explain the observed offset between the small hydrocarbons 
%and H$_2$ emission (Pety et al. 2005a).  
The exact density values still depend on the 
assumed cloud depth  and temperatures.
In any case, the inferred \textit{shell} of dense molecular gas has  
high thermal pressures $\sim$(5--10)$\times$10$^6$~K~cm$^{-3}$ 
and this can be the signature of the processes driving the slow expansion of the
PDR. Therefore, the most shielded clumps undergo effective line cooling 
and  the regions of lower density should  be compressed due to their lower internal pressure.
Recent hydrodynamical simulations of  the expansion of ionization and dissociation fronts
 around massive stars also predict that a high density molecular
\textit{shell} (10--100 times the ambient density) will be swept up behind  the ionization 
front (Hosokawa \& Inutsuka 2005 a\&b). The density, pressure and temperature profiles and 
values
predicted by these simulations at $\sim$0.5~Myr (the Horsehead formation timescale derived 
from its velocity gradients by Pound et al. 2003 and Hily-Blant et al. 2005) qualitatively 
reproduce the values inferred from our molecular line observations and modeling.
Hence, a shock front driven by the expansion of the ionized gas is probably compressing 
the cloud edge to the high densities 
observationally inferred in this work. Specific hydrodynamical simulations for the 
particular source physical conditions and comparison with observations will be appreciated.
As noted by Hosokawa \& Inutsuka (2005), the dynamical expansion of a HII region, PDR and 
molecular \textit{shell} in a cloud with a density gradient has not been studied well. 
We suggest the Horsehead PDR as a good target.

\subsection{Temperatures}

Molecular excitation, radiative transfer and chemical models are used to 
derive realistic abundances. The gas temperature impacts many aspects of these 
computations (e.g. chemical reaction rates and collisional excitation), and thus, the
density and abundance uncertainties also reflects our incomplete knowledge of the 
thermal structure. The problem is not straightforward, since a steep temperature 
gradient is also expected in PDRs, and also because the most appropriate tracers of the warm gas
lie at higher frequencies. The Horsehead PDR may not be an extreme
case, since its FUV radiation field is not very high and photoelectric heating
alone will not heat the gas to high temperatures as long as the gas
is FUV--shielded. Nevertheless, our thermal balance calculations quickly lead
to T$_k$ $\simeq$10~K. According to observations, this temperature is too
low, especially in the first $\delta x \sim$30$''$ representing the PDR edge. In this
work we have (observationally) adopted T$_k$=30~K as the minimum gas 
temperature in our PDR calculations. In forthcoming
works we will concentrate on the thermal structure of the PDR. Here we
only note that either the cooling is not so effective and/or  extra heating mechanisms 
need to be considered.
%Since an homogeneous unidimensional cloud only illuminated by one side
%may not be the most exact description of the PDR we also run some
%photochemical models illuminating the cloud also in the other side
%with a $\chi$=1 radiation field. This modification resulted only in
%a increase of $\Delta T_k$ $\sim$2~K in the shielded regions.
The cosmic ray ionization rate was also increased by a factor 
$\sim$5 but it only modifies the gas temperature by $\Delta T_k$ $\sim$4~K.

Under these circumstances we have to conclude that the gas, or at least
a fraction of it, is likely to be warmer than predicted.
We note that this problem is not new. Again, high--$J$ $^{13}$CO and NH$_3$ observations 
have previously probed the existence of warm gas ($\sim$150~K) in regions where
standard heating mechanisms fail to predict those values 
(Graf et al. 1990; Batrla \& Wilson 2003). 
More recently, Falgarone et al. (2005)  reported ISO observations of H$_2$ in the lowest
five pure rotational lines S(4) to S(0) (8~$\mu$m to 28~$\mu$m) toward diffuse ISM gas.
The observed S(1)/S(0) and S(2)/S(0) line ratios are too large to be compatible with the
 PDR models. These authors suggested that MHD shocks 
(Flower \& Pineau des Forets 1998) or magnetized vortices,
which are natural dissipative structures of the intermittent dissipation of turbulence
(Joulain et al 1998),
locally heat the gas at temperatures up to $\approx 1000$~K. These structures
add two heating sources: \textit{i)} viscous heating through large velocity
shear localized in tiny regions and \textit{ii)} ion-neutral drift heating due
to the presence of magnetic fields (ambipolar diffusion).  As shown by Falgarone
et al (2006), these dissipative structures are also able to trigger a
hot chemistry, that is not accessible to models that do not take into account the
gas dynamics. These results suggest that additional mechanical heating processes are at work.
 We  propose that the shock waves driven by the expansion of the HII region and PDR
compress the molecular gas in the cloud edge and provide it with an additional heating source.

\subsection{CS and HCS$^+$ chemistry}

According to the last (but not least) molecular data affecting CS chemistry and excitation,
 the mean CS abundance in the Horsehead PDR, $\chi(CS)$=(7$\pm$3)$\times$10$^{-9}$,
implies a gas sulfur abundance  of S/H $\sim$(3.5$\pm$1.5)$\times$10$^{-6}$, only a 
factor $\lesssim$4 smaller than the solar value (\cite{asp05}). 
Even  lower sulfur depletion values are possible if the gas is significantly warmer
than 30~K.
Thus, the gas  phase sulfur abundance  is very close to the
undepleted value observed in the diffuse ISM and not to the depleted value invoked
in dense molecular clouds  (e.g. Millar \& Herbst 1990). However,
the observed CS/HCS$^+$ $\simeq$175 abundance ratio can only be reproduced
by photochemical models by considering the HCS$^+$ peak abundance,
otherwise, larger ratios ($\sim$10$^3$) are predicted.
Therefore, either the observed
HCS$^+$ only traces the surface of the cloud where its abundance peaks, or
chemical models underestimate the HCS$^+$ production rate.
%The latter is a consequence of the almost constant $\chi(HCS^+)$ predicted
%value for S/H $>$2$\times$10$^{-6}$.
In any case, the predicted CS/HCS$^+$ abundance ratio scales with the  gas phase 
sulfur abundance. The largest ratios are expected at the lowest sulfur depletions.
However, the observed CS/HCS$^+$ $\sim$10 ratio in the diffuse ISM
(Lucas \& Listz 2002) is even lower than in the Horsehead.
Thus, we have to conclude that present chemical models still fail to reproduce
the observed CS/HCS$^+$ abundance ratio, at least in the stationary regime.
Time--dependent photochemical computations may also help to understand
the dynamical expansion of the dissociation front and the evolving
molecular abundances.
Besides, Gerin et al. (1997) noted that larger HCS$^+$ abundances are expected
if the gas is in a high ionization phase. We have computed that if the cosmic ray ionization
rate is increased by a factor of $\sim$5, the predicted HCS$^+$ abundance inside
the cloud (A$_V$=10~mag) interestingly matches our inferred value and the
CS/HCS$^+$  abundance ratio gets much closer to the observed ratio without the need 
of  taking the HCS$^+$ abundance peak.

For the physical and FUV illuminating conditions prevailing in the Horsehead
PDR, most of the gas phase sulfur is locked in S$^+$ for A$_V$ $\lesssim$2~mag
and $\chi(HCS^+)$ reaches its maximum value. 
Besides, the derived gas phase sulfur abundance
is large enough to keep $\chi(e^-)$$>$10$^{-7}$ for A$_V$$\lesssim$3.5~mag.
HCS$^+$ and
S\,{\sc ii} recombination lines trace the skin of molecular clouds
where S$^+$ is still the dominant form of sulfur.
In the scenario proposed by Ruffle et al. (1999),
these S$^+$ layers will be responsible of the sulfur depletion due to
more efficient sticking collisions on negatively charged dust grains than
in the case of neutral atoms such as oxygen.
Even in these  regions, still dominated by
photodissociation, CS and HCS$^+$ abundances are quickly enhanced 
compared to other sulfur molecules. In fact, we predict that
CS is the most abundant S--bearing molecule in the external layers
where S$^+$ is still more abundant than neutral sulfur.
These results are consistent 
with our PdBI detection of CS close to the PDR edge and show that
CS is a PDR tracer. These findings are  consistent with
observations of S--bearing species in the diffuse ISM where
CS is more abundant than SO$_2$, H$_2$S and SO (Lucas \& Listz 2002). 

Between A$_V$ $\sim$2 and $\sim$8~mag the  S$^+$ abundance smoothly decreases.
Since S$^+$ is a good source of electrons, the electronic fractionation
also decreases accordingly. 
HCS$^+$, and thus CS, present an abundance minimum in these
layers. Neutral atomic sulfur is now the most abundant
S--bearing species. Therefore, observations of the [S\,{\sc i}]25~$\mu$m
fine structure line will basically trace these intermediate
layers of gas where S--bearing molecules have not reached their abundance peak.
However, the exctitation energy of the [S\,{\sc i}]25~$\mu$m line
(the upper level energy is  $\sim$570~K) is too high compared to the thermal energy
available in the  regions where the neutral sulfur abundance peaks  ($T_k$~$\simeq$30~K)
and no detectable emission is expected.
In fact, no Spitzer/IRS  line detection has
been reported in the Horsehead (L.~Verstraete, private com.).
However, since most of the neutral atomic sulfur will remain in the ground--state, 
the presence of a background IR source (e.g. in face--on PDRs)
may allow, with enough spectral resolution and continuum sensitivity, 
the detection of the [S\,{\sc i}]25~$\mu$m line in absorption.

On the other hand, sulfur in diffuse ionized gas outside the molecular cloud is in
the form of  sulfur ions. Mid--IR [S\,{\sc iii}] fine structure 
lines have been detected around the Horsehead with IRS/Spitzer 
(L. Verstraete, private com.).
In the shielded gas, sulfur is mostly locked in
S--bearing molecules together with a smaller fraction in atomic form. 
Our models predict that species such as SO will be particularly abundant.
Jansen et al. (1995) also noted that the low gas phase sulfur abundance needed to explain the
CS abundance in the Orion~Bar PDR was incompatible with the observed 
high H$_2$S/CS$\sim$0.5 
abundance ratio. Therefore, a complete understanding of the sulfur chemistry will
only be achieved when all the major sulfur molecules can be explained. In a forthcoming paper we
analyse the photochemistry, excitation and radiative transfer of several S-bearing
molecules detected by us in the Horsehead PDR.

%It is also an open question if large quantities of sulfur are
%in S--bearing molecules not detected yet.
%We have currently started a
%study of several S--bearing molecules in the Horsehead PDR to go deeper
%on the analysis of the gas phase sulfur chemistry.
%
%  S MOLECULES
%

%\clearpage

\section{Summary and Conclusion}

We have presented interferometric maps of the Horsehead PDR in the CS $J$=2--1 line 
at a $3.65'' \times 3.34''$  resolution together with single--dish observations of several
rotational lines of CS, C$^{34}$S and HCS$^+$. We have studied the CS photochemistry,
excitation and radiative transfer using the latest  HCS$^+$ and OCS$^+$ dissociative
recombination rates  (Montaigne et al. 2005) and CS collisional cross--sections 
(Lique et al. 2006). The main conclusions of this work are as follows:

\begin{enumerate}

\item CS and C$^{34}$S rotational line emission reveals
mean densities around $n(H_2$)=(0.5--1.0)$\times$10$^{5}$~cm$^{-3}$.
The CS $J$=5--4 lines show narrower line widths than the low--$J$ CS lines  and  
require higher density gas components, $\sim$(2--6)$\times$10$^{5}$~cm$^{-3}$, not  
resolved by  a  $\sim$10$''$ beam.
These values are larger than previous estimates based on CO observations.
It is likely that  clumpiness at scales below $\sim$0.01~pc and/or a low density 
envelope play a role in the CS line profile formation.

%Observation of other high dipole moment molecules will be needed 
%to better characterize the density inhomogeneity or possible clumpyness
%of the medium.

\item Nonlocal, non--LTE radiative transfer models of optically thick CS lines
and optically thin C$^{34}$S lines provide an accurate determination
of the CS abundance, $\chi(CS)$=(7$\pm$3)$\times$10$^{-9}$.
We show that radiative transfer and opacity effects play a role
in the resulting CS line profiles but not in C$^{34}$S lines. 
Assuming the same physical conditions for the HCS$^+$ molecular ion,
 we find $\chi(HCS^{+})$=(4$\pm$2)$\times$10$^{-11}$.

\item According to photochemical models, the gas phase sulfur abundance 
required to reproduce these CS and HCS$^+$ abundances is 
S/H=(3.5$\pm$1.5)$\times$10$^{-6}$, only a factor $\sim$4 less abundant than the
solar elemental abundance.
Larger sulfur  abundances are possible if the gas is significantly warmer.
Thus, the sulfur abundance in the PDR is very close to the
undepleted value observed in the diffuse ISM.
The predicted CS/HCS$^+$ abundance ratio is  close to the observed
value of $\sim$175, especially if predicted HCS$^+$ peak abundances
are considered. If not,  the HCS$^+$ production is  underestimated unless 
the gas is in a higher ionization phase, e.g.
if the cosmic ray ionization rate is increased by  $\sim$5.
%A full inventory of S--bearing molecules in the PDR is needed
%to test this and other limitations of the sulfur chemical network.

\item High angular resolution PdBI maps reveal that the CS  emission
does not follow the same morphology shown by the small hydrocarbons emission
the PDR edge. %In order to analyze the aperture synthesis observations,
%We have developed a  nonlocal non--LTE  radiative transfer code including 
%dust and cosmic background illumination for an edge--on plane--parallel 
%cloud with possible inclination with respect to the line of sight. 
In combination with 
previous PdBI C$^{18}$O  observations we have modeled the PDR edge and confirmed that a
 steep density gradient is needed to reproduce CS and C$^{18}$O
observations.  The resulting density profile qualitatively agrees to that predicted  in 
numerical simulations of a shock front compressing the PDR edge to high 
densities, $n(H_2$)$\simeq$10$^{5}$~cm$^{-3}$, and
high thermal pressures, $\simeq$(5--10)$\times$10$^6$~K~cm$^{-3}$.

\item Conventional PDR heating and cooling mechanisms fail to reproduce
the temperature of the warm gas observed in the region  by at
least a factor $\sim$2. Additional mechanical heating mechanisms associated with the gas 
dynamics may be needed to account for the warm gas. The thermal structure of the PDR
edge is still not fully constrained from observations. This fact
adds uncertainty to the abundances predicted by photochemical models.

\end{enumerate}

We have shown that many physical and chemical variations
in the PDR edge occur at small angular scales. In addition, the molecular
inventory as a function of the distance from the illuminating source
can only be obtained from millimeter interferometric observations.
High angular resolution observations contain detailed information
about density, temperature, abundance and structure of the cloud,
but only detailed radiative transfer and photochemical models for each
given source are able to extract
the information. A minimum description of the source geometry is
usually needed. Future observations with ALMA will allow us to characterize
in much more details many energetic surfaces such as PDRs.

\begin{acknowledgements}
We are grateful to the \IRAM{} staff at Plateau de Bure, Grenoble and
Pico Veleta for the remote observing capabilities and  competent help with 
the observations and data reduction.
We also thank BASECOL, for the quality of data and information provided, 
and  F.~Lique for sending us the CS collisional rates prior to publication.
JRG thanks J.~Cernicharo, F.~Daniel and I. Jim\'enez-Serra for fruitful discussions. 
We finally thank John Black, our referee, for useful and encouraging comments.
JRG was supported by the french \textit{Direction de la Recherche} 
and by a \textit{Marie Curie Intra-European Individual
Fellowship} within the 6th European Community Framework Programme,
contract MEIF-CT-2005-515340.
\end{acknowledgements}

\clearpage

\begin{appendix}
\section{Nonlocal, non--LTE radiative transfer}

Radiative transfer in a medium dominated by gas phase molecules
and dust grains requires the solution of the radiative transport (RT) equation
for the radiation field together with the equations governing the relative
level populations of the considered species. In the case of  rotational
line emission (far--IR to mm domain),
scattering from dust grains can be usually neglected from the RT equation
and steady state statistical equilibrium can be assumed for  molecular
populations. However, physical conditions in ISM clouds are such
that molecular excitation is usually far from LTE. Therefore, a minimum treatment
of the nonlocal coupling between line+continuum radiation and level
populations is required. In this appendix we describe in more detail
the simple model developed for this work.

\subsection{Monte Carlo methodology for gas and dust
radiative transfer in plane-parallel geometry}
\label{sub-apendix1}

The Monte Carlo methodology or its modifications is a simple and widely
adopted approach when one has to deal with moderately thick lines 
and flexibility to explore different geometries is required
(see van Zadelhoff et al. 2002 and references therein).
In this work, the $\textit{classical}$ description of the Monte Carlo 
approach for non--LTE line transfer (Bernes 1979) has been extended 
to include the dust emission/absorption and their effect on the source function.
The code was originally developed in \textit{fortran90}
for spherical symmetry (Goicoechea 2003) and has been enlarged to 
semi--infinite plane--parallel geometry (from face-- to edge-on). 
Thus, numerical discretization is transformed from spherical shells to slabs. 
The model includes illumination from the cosmic background at both surfaces.

\FigPlgeo

The variation of the radiation field intensity along any photon path $s$
is related to the emission and absorption properties of the medium
(scattering neglected) through 
\begin{equation}
\frac{dI_{\nu}}{ds}= j_{\nu}-\alpha_{\nu}I_{\nu} 
\label{eq-ETR1}
\end{equation}   
where  $\alpha_{\nu}$ [cm$^{-1}$] and 
$j_{\nu}$ [erg s$^{-1}$ cm$^{-3}$ Hz$^{-1}$ sr$^{-1}$] are
the total (gas+dust) absorption and emissivity coefficients at a 
given frequency $\nu$. The normal path to any plane--parallel slab is 
thus $dz=\mu~ds$, with $\mu = cos\,\theta$ and where $\theta$ is
the angle between $z$ and $s$
(see Fig.~\ref{fig:_plgeo}). Equation \ref{eq-ETR1}
is thus  written as
\begin{equation}
\mu\,\frac{dI_{\nu}}{d\tau}= S_{\nu}-I_{\nu} 
\label{eq-ETR2}
\end{equation}   
where the differential optical depth is given by gas and dust contributions,
$d\tau = \alpha_{\nu}~dz$, and $S_\nu = j_{\nu}/\alpha_{\nu} $ is
referred to as the source function. Continuum emissivity from dust
is assumed to be thermal and given by
\begin{equation}
j_{\nu}^{\,dust}= \alpha_{\nu}^{\,dust}B_{\nu}(T_d)  
\label{eq-dust_emi}
\end{equation}   
where $B_{\nu}$ is the Planck function at a given dust
temperature, T$_d$, and  $\alpha_{\nu}^{\,dust}$ is computed from any of 
the dust mass absorption cross--sections available in the literature
(e.g. Draine \& Lee, 1984, Ossenkopf and Henning, 1994). For practical
purposes, the dust absorption coefficient is assumed to be constant
in all the passband around each line frequency. Hence, the number
of dust continuum photons emitted per second in a given cell of material is
$(4\pi/hc)\,j_{\nu}^{\,dust}\,V_m\,\Delta$v where $V_m$ is the cell volume
and $\Delta$v the considered passband in velocity units.
Although the inclusion of dust almost does not affect molecular excitation in 
our work, it is included for consistency
and for making predictions of higher frequency lines where it has larger effects. 

Molecular lines occur at discrete frequencies, $\nu_{ij}$,
where $i$ and $j$ refer to upper and lower energy levels with
$n_i$ and $n_j$ [cm$^{-3}$] populations respectively.
Gas emission and absorption coefficients, as a function of velocity,
are defined as
\begin{equation}
j_{\nu}^{\,gas} = \frac{hc}{4\pi}\,n_i\,A_{ij}\,\phi 
\,\,\,\,\,\,\,\,\,;\,\,\,\,\,\,\,\,\, 
\alpha_{\nu}^{\,gas} = \frac{hc}{4\pi}\,(n_j\,B_{ji}-n_i\,B_{ij})\,\phi
\label{eq-gas_emi}
\end{equation}   
where $B_{ji}$, $B_{ij}$, and $A_{ij}$  are the 
transition probabilities, or Einstein coefficients, for absorption, and for
induced and spontaneous emission respectively. We have assumed
the same line Doppler profile (in velocity units)
for emission and absorption
\begin{equation}
\phi = \frac{1}{b\sqrt \pi}\exp\left(-\frac{\bf{v} + \bf{v_f}\cdot\bf{s}}{b}\right)^2
\label{eq-profile}
\end{equation}   
and thus considered Gaussian Doppler microturbulent and thermal broadening 
characterized by the broadening parameter $b^2=$ v$_{turb}^{2}$+ v$_{th}^{2}$. 
Note that any arbitrary velocity field $\bf{v_f}$ can be included.
Here we take the possibility of having a velocity field normal
to the slabs, i.e. $\bf{v_f}$=v$_f$\,(z).

Generally speaking, the relative level populations of a considered 
molecule $m$ are determined by collisions with other molecules, 
and/or by radiative effects caused by the  cosmic background 
and/or by the dust continuum emission. The particular physical conditions,
type of molecule and spectral domain will determine the dominant processes
through the steady state statistical equilibrium equations
\begin{equation}
n_i \sum_{j\neq i} [R_{ij} + C_{ij}] = \sum_{j\neq i} n_j [R_{ji} + C_{ji}]
\,\,\,\,\,;\,\,\,\,\, n_{tot} = \sum_{J=1}^{rot\,levels} n_J 
\label{eq-ec-estad1}
\end{equation}  
where $C_{ij}$ and $R_{ij}$ [s$^{-1}$] are
the collisional and radiative transition rates between
$i$ and $j$ levels. For the collisional rates of species $m$ 
(CS, C$^{34}$S, C$^{18}$O and HCS$^+$) we have considered 
\begin{equation}
C_{ij} = \gamma_{ij}^{m}\,(H_2)\;n(H_2)+\gamma_{ij}^{m}\,(He)\;n(He)
+\gamma_{ij}^{m}\,(H)\;n(H)
\label{eq-ec-prob_col1}
\end{equation}  
where $\gamma_{ij}^{m}$ [cm$^{3}$\,s$^{-1}$] are the temperature
dependent collisional de--excitation rate coefficients of $m$ with 
 collisional partners H$_2$ or He. If unknown,
excitation rate coefficients are computed through detailed balance.
For consistency with the PDR modeling 
we have estimated the collisional rates with H atoms
(simply by scaling from the He rates), since H may
be the dominant partner in the  diffuse regions.
Radiative rates are 
\begin{equation}
R_{ij} = A_{ij} + B_{ij} \bar{J}_{ij}\,\,\,\,\,\,\,;\,\,\,\,\,\,\,
R_{ji} = B_{ji} \bar{J}_{ji}
\label{eq-ec-prob_rad1}
\end{equation}
where $\bar{J}_{ij}$ is the intensity of the radiation field  integrated 
over solid angles and over the line profile. External illumination
by cosmic background, dust continuum emission and line photons from
different spatial regions contribute to $\bar{J}_{ij}$.
Hence, the solution of molecular excitation implicitly depends
on the nonlocal radiation field, which obviously depends on 
level populations in many cloud points.
$\bar{J}_{ij}$ is explicitly computed in the Monte Carlo approach,
and thus, the RT-excitation problem is solved iteratively until desired
convergence in some physical parameter (generally the level populations) 
is achieved. LTE level populations at a constant fictitious  T$_{ex}$ 
were used for the first iteration. In the case of CS modeling,
T$_{rot}$ from the observational rotational--diagrams 
(Fig.\ref{fig:ROTdiag}) was used.

\FigCSbench

The RT problem is then simulated by the emission of a determined number
of model photons (both sides external illumination, continuum and line
photons) in a similar way to that originally described by Bernes (1979).
Model photons represent a large quantity of $real$ photons
randomly distributed over the line profile \ref{eq-profile} and
emitted at random cloud positions and directions.  
Each model photon is followed through the different slabs until
it escapes the cloud or until the number of represented real photons 
become insignificant. Note that the angle $\theta$ between the 
photon direction and the normal to the slabs remains constant in all 
the photon path. In spherical geometry the $\theta$ angle between the 
photon direction and the radial direction changes in each photon step 
and thus has to be computed repeatedly. 
In addition, model photons sent in the $cos\,\theta\simeq0$ direction
in semi--infinite plane--parallel geometry will almost never escape
the cloud. Thus, a minimum number of represented real photons
is defined otherwise the photon is not followed anymore.
In this way, the Monte Carlo simulation explicitly computes the 
induced emissions caused by the different types of model photons 
in all the slabs. At the end of the simulation, an averaged value 
for the $B_{ij} \bar{J}_{ij}$ that independently accounts for external 
illumination, continuum emission and line emission is stored for every 
slab ($\sum S_{ij,m}$ in Bernes formalism).
%The knowledge of these values allows a detailed analysis of the molecular
%excitation across the cloud and helps to evaluate the dominant contributions
%to radiative effects at any cloud location. 
Populations are then  obtained in each slab by solving:
\begin{equation}
n_{i}\sum_{j\neq i}[A_{ij}+\sum S_{ij,m}+C_{ij}]= \sum_{j\neq i}n_j\, 
[\frac{g_i}{g_j} \sum S_{ij,m}+C_{ji}] 
\label{eq-eq_stat2_mc}
\end{equation}
A reference field for all types of model photons was included to reduce
the inherent random fluctuations (i.e. the variance) of any Monte Carlo 
simulation (see Bernes 1979, for details).
When a prescribed convergence in level the populations is reached,
the total (gas+dust) source function is completely determined and the
emergent intensity can be easily computed by integrating Eq.~\ref{eq-ETR2}.

For spherical geometry, the code was successfully 
benchmarked against two test problems, the \textit{Bernes' CO cloud}
(Bernes 1979) and the \textit{HCO$^+$ collapsing cloud, problem 2a} 
of the 1999 Leiden benchmark (van Zadelhoff et al. 2002).
In the case of plane--parallel geometry, several thermalization tests
 for the CS excitation (without dust emission) were successfully performed.
Excitation temperatures as a function of the normal 
coordinate to slabs $z$ are shown for CS $J$=1--0, 2--1, 3--2 and 5--4 transitions
in Fig.~\ref{fig:CS_bench}.
Model parameters are T$_k$=20~K, $n$(H$_2$)=10$^5$~cm$^{-3}$ 
and $\chi$(CS)=7$\times$10$^{-9}$. Different excitation conditions
are considered. Upper model: $A_{ij}/C_{ij}=0$ and T$_{ex}$ is correctly
thermalized  to T$_{kin}$ (LTE). Middle model: collisional rates from
Lique et al., (filled squares; 2006) and Turner et al. (empty circles; 1992) 
and resulting non--LTE excitation (see section~\ref{CS_exc}). As noted
by Lique et al., their new collisional rates produce
larger excitation temperatures, especially as $J$ increases. For the Horsehead
physical conditions this implies that the estimated densities and/or
abundances with Turner et al. collisional rates are $\sim$10$\%$ larger.
Lower model: collisional excitation neglected and 
T$_{ex}$ is radiatively thermalized to the cosmic background temperature 
at T$_{bg}$=2.7~K.

\FigModelgeo{}

\subsection{A model for an edge--on cloud with inclination}
\label{sub-apendix2}

In order to benchmark the spatial distribution of the  PDR code abundance
predictions with our interferometric line observations, we now can use the simple  model
described above to compute the synthetic spectrum of a required molecule.
%The Horsehead PDR geometry in the plane of the sky and its illumination 
%conditions are such that, as a first approximation, it can be ideally 
%modeled as an almost edge--on plane-parallel source with a possible small 
%angle of inclination $\varphi$ respect to the line of sight.
To do that, the PDR code output was used as an input for the RT calculation.
In particular, the density profile (both $n(H_2)$ and $n(H)$), temperature
profile (both T$_k$ and T$_d$) and species abundance are 
carefully interpolated from the PDR spatial grid output. In practice, 
the slab discretization for the RT calculation has to be precise enough 
to sample the abundance, density and temperature variations at the edge 
of the PDR (where most of the changes occur). In most RT computations, 
$\sim$50~slabs were judged to give satisfactory sampling of the PDR variations.
For an edge--on configuration, after a Monte Carlo simulation,  RT 
equation \ref{eq-ETR2} was integrated in a grid of different lines
of sight (similar to impact parameters in spherical geometry) as depicted 
in Fig.~\ref{fig:geometry}. Lines of sight can be inclined by an
angle $\varphi$ respect to the slabs normal ($ds = dz/sin\;\varphi$).  
Therefore, the maximum integration path is $l_{depth}/cos\;\varphi$
where $l_{depth}$ is the assumed cloud spatial depth.
To produce a synthetic map, results are then convolved in a grid of cloud points
with an angular resolution characterized by a gaussian with  hpbw
equal to that of the synthesized interferometric beam.

\end{appendix}


\begin{thebibliography}{}



\bibitem[Abergel et al. 2002]{ab02} 
Abergel, A., Bernard, J.~P., Boulanger, F. et al. 2002,
A\&A, 389, 239.

\bibitem[Abergel et al. 2003]{ab03} 
Abergel, A. et al. 2003,
A\&A, 410,  577.

%\bibitem[Anders \& Grevesse 1989]{ag89} 
%Anders, E. \& Grevesse, N. 
%1989, GeCoA, 53, 197-214

\bibitem[Anthony-Twarog 1982]{at82}
Anthony-Twarog, B.~J. 1982,
A\&J,  87, 1213.
 
\bibitem[Asplund et al. 2005]{asp05}
Asplund, M.,  Grevesse, N.,  \& Sauval, A. J. 2005
in ASP Conf. Ser. 336, Cosmic Abundances as Records of Stellar Evolution and Nucleosynthesis, ed. F. N. Bash \& T. G. Barnes, 25 

\bibitem[Batrla \& Wilson 2003]{bat03}
Batrla, W. \& Wilson, T. L.
2003, A\&A, 408, 231


\bibitem[Bernes 1979]{ber79}
Bernes, C. 1979, A\&A, 73, 67.

\bibitem[Bogey et al. 1981]{Bog81}
Bogey, M., Demuynck, C., \& Destombes, J. L.
1981, Chem.Phys.Letters, 81, 256

\bibitem[Boogert et al. 2000]{bog00}
Boogert, A. C. A. et al.
2000, A\&A, 360, 683.

\bibitem[Cernicharo \& Guelin 1987]{cer87}
Cernicharo, J. \& Guelin, M. 
1987, A\&A, 176, 299.

\bibitem[Dartois 2005]{dar05} 
Dartois, E. 2005, Space Science Reviews, 119, Issue 1-4, p. 293-310

\bibitem[Draine 1978]{dra78}
Draine, B. T.  1978,
ApJS, 36, 595.

\bibitem[Draine \& Lee 1984]{dra84}
Draine, B. T.  \& Lee, H.M.
1984, ApJ, 285, 89.

\bibitem[Emerson \& Graeve 1888]{eme88}
Emerson, D. T., \& Graeve, R.
1988, A\&A, 190, 353.

\bibitem[Falgarone et al.  2005]{fal05}
Falgarone, E.,  Verstraete, L.,  Pineau Des Forets, G. \& Hily-Blant, P.
2005,  A\&A, 433, 997.

\bibitem[Falgarone et al.  2006]{fal06}
Falgarone, E., Pineau Des Forets, G., Hily-Blant, P. \& Schilke, P.
2006,  A\&A, in press.

\bibitem[Flower \& Pineau des Forets 1998]{flo98}
Flower, D. R., \& Pineau des Forets, G. 
1998,  MNRAS, 297, 1182.

\bibitem[Flower 2001]{flo01}
Flower, D. R.
2001, JPhB, 34, 2731

\bibitem[Frerking et al. 1980]{fre80}
Frerking, M. A., Wilson, R. W., Linke, R. A., \& Wannier, P. G.
1980, ApJ, 240, 65.


\bibitem[Garc\'{\i}a--Rojas  et al. 2006]{gar06}
Garc\'{\i}a-Rojas, J., Esteban, C., Peimbert, M., Costado, M. T., Rodríguez, M., 
Peimbert, A. \& Ruiz, M. T.
2006, MNRAS, 368, 253

\bibitem[Gerin et al. 1997]{ger97}
Gerin, M. et al. 
1997, A\&A, 318, 579.


\bibitem[Gibb et al. 2004]{gi04}
Gibb, E. L., Whittet, D. C. B., Boogert, A. C. A., \& Tielens, A.G.G.M.
2004, ApJS, 151, 35.

\bibitem[Goicoechea 2003]{goi03}
Goicoechea, J.R.
2003, Ph. D. Thesis, Universidad Autonoma de Madrid, September 2003.

\bibitem[Goldsmith \& Langer 1999]{gol99}
Goldsmith, P.F.,  \& Langer, W.D.
1999, ApJ, 517, 209.

\bibitem[Gonzalez-Alfonso \& Cernicharo 1993]{gon93}
Gonz\'alez-Alfonso, E. \& Cernicharo, J.
1993, A\&A, 279, 506.

\bibitem[Graedel et al. 1982]{gra82}
Graedel, T. E., Langer, W. D., \&  Frerking, M. A.
1982, ApJS, 48, 321.

\bibitem[Graf et al. 1990]{gra90}
Graf, U. U., Genzel, R., Harris, A. I., Hills, R. E., Russell, A. P. G. \& Stutzki, J.
1990, ApJ, 358, L49.

\bibitem[Gueth et al. 1996]{gu96}
Gueth, F., Guilloteau, S. \& Bachiller, R.
1996, A\&A, 307, 891-897.

\bibitem[Habart et al. 2005]{ha05}
Habart, E., Abergel, A., Walmsley, C. M., Teyssier, D. \& Pety, J.
2005, A\&A, 437, 177-188.

\bibitem[Hily-Blant et al. 2005]{hil05}
Hily-Blant, P., Teyssier, D., Philipp, S. \& Gusten, R.
2005, A\&A, 440, 909.

\bibitem[Hosokawa \& Inutsuka]{hos05a}
Hosokawa, T. \& Inutsuka S.
2005a, ApJ, 623, 917.

\bibitem[Hosokawa \& Inutsuka]{hos05b}
Hosokawa, T. \& Inutsuka S. 
2005b, astro-ph/0511165

\bibitem[Howk et al. 2006]{how06}
Howk, J. C., Sembach, K. R.,  \& Savage, B. D. 2006
2006, ApJ, 637, 333.

\bibitem[Irvine et al. 1985]{ir85}
Irvine, W. M., Schloerb, F. P., Hjalmarson, A. \& Herbst, E.
1985, Protostars and planets II (A86-12626 03-90). 
Tucson, AZ, University of Arizona Press, 1985, 579-620

\bibitem[Jensen et al. 1995]{jan95}
Jansen, D. J., Spaans, M., Hogerheijde, M. R., \& van Dishoeck, E. F.
1995, A\&A, 303, 541.

\bibitem[Joulain et al. 1998]{jou98}
Joulain, K., Falgarone, E.,  Pineau des Forets, G.  \&  Flower, D.
1998, A\&A, 340, 241.

\bibitem[Le Bourlot et al. 1993]{jlb93}
Le Bourlot, J., Pineau Des Forets, G.,  Roueff, E.,  \& Flower, D. R.
1993, A\&A, 267, 233.

\bibitem[Le Petit et al. 2006]{lp06}
Le Petit, F., Nehm\'e, C, Le Bourlot, J. \& Roueff, E.
2005, ApJS, preprint, doi:10.10861'503252'.

%\bibitem[Lehner et al. 2004]{leh04}
%Lehner, N.,  Wakker, B. P. \& Savage, B. D.
%2004, ApJ, L615, 767.

\bibitem[Lique et al. 2006]{liq06}
Lique, F., Spielfiedel, A. \& Cernicharo, J.
2006, A\&A, 451, 1125

\bibitem[Lis \& Schilke 2003]{ls03}
Lis, D. C., \& Schilke, P.
2003, ApJ, 597, L145.

\bibitem[Lucas \& Listz 1998]{ll98}
Lucas, R., \& Liszt, H.
1998, A\&A, 337, 246.

\bibitem[Lucas \& Listz 2002]{ll02}
Lucas, R., \& Liszt, H.
2002, A\&A, 384, 1054.

\bibitem[Mart\'{\i}n-Hern\'andez  et al. 2002]{mar02}
Mart\'{\i}n-Hern\'andez et al. 2002
2002, A\&A, 381, 606

\bibitem[Mart\'{\i}n et al. 2005]{mar05}
Mart\'{\i}n, S., Mart\'{\i}n-Pintado, J., Mauersberger, R., Henkel, C., \& Garc\'{\i}a-Burillo, S.
2005, ApJ, 620, 210

\bibitem[Millar \& Herbst 1990]{mil90}
Millar, T. J. \& Herbst, E. 
1990, A\&A, 231, 466.

\bibitem[Montaigne et al. 2005]{mon05}
Montaigne, H., Geppert, W. D., Semaniak, J., et al.
2005, ApJ, 631, 653.

\bibitem[Monteiro 1984]{mon84}
Monteiro, T.
1984, MNRAS, 210, 1


\bibitem[Neufeld et al. 2005]{neu05}
Neufeld, D.A., Wolfire, M.G., \& Schilke, P.
2005, ApJ, 628, 60.


\bibitem[Ohishi \& Kaifu 1998]{ohi98}
Ohishi, M. \& Kaifu, N.
1998,  Chemistry and Physics of Molecules and Grains in Space. Faraday Discussions No. 109. 
The Faraday Division of the Royal Society of Chemistry, London, 1998, 205

\bibitem[Ossenkopf \& Henning 1994]{oss94}
Ossenkopf, V. \&  Henning, Th. 
1994, A\&A, 291, 943.

\bibitem[Palumbo et al. 1997]{pal97}
Palumbo, M.E., Geballe, T.R., Tielens, A.G.G.M.
1997, ApJ, 479, 839.

\bibitem[Pankonin \& Walmsley 1978]{pan78}
Pankonin, V., \& Walmsley, C.M. 
1978, A\&A, 64, 333.

\bibitem[Penzias \& Burrus 1973]{pen73}
1973, ARA\&A, 11, 51.

\bibitem[Pety et al. 2005a]{pety05} 
Pety, J., Teyssier, D., Foss\'e, D., Gerin, M., Roueff, E., Abergel, A., 
Habart, E. \&  Cernicharo, J. 2005,
A\&A, 435, 885-899. 

\bibitem[b]{pet05b} 
Pety, J. 2005, in SF2A-2005: Semaine de l'Astrophysique Francaise, 721

\bibitem[Pound et al. 2003]{prb03}
Pound, M.~W., Reipurth, B. \&  Bally, J. 
2003, A\&J, 125, 2108-2122.

\bibitem[Reipurth \& Bouchet 1984]{rei84}
Reipurth, B., \& Bouchet, P.
1984, A\&A, 137, 1.

\bibitem[Ruffle et al. 1999]{ruf99}
Ruffle, D. P., Hartquist, T. W., Caselli, P. \& Williams, D. A.
1999, MNRAS, 306, 691.

%\bibitem[Sofia et al. 1994]{sof94}
%Sofia, U.J., Cardelli, J.A. \& Savage, B.D.
%994, ApJ, 430, 650.

\bibitem[Tieftrunk et al. 1994]{tie94}
Tieftrunk, A., Pineau des Forets, G., Schilke, P. \& Walmsley, C. M.
1994, A\&A, 289, 579.


\bibitem[Teyssier et al. 2004]{tf04} 
Teyssier, D., Foss\' e, D., Gerin, M., Pety, J.,  Abergel, A. \& Roueff, E.
2004, A\&A, 417, 135-149.

\bibitem[Turner et al. 1992]{tur92}
Turner, B. E., Chan, K.W, Green, S., \& Lubowich, D. A.
1992, ApJ, 399, 114.

%\bibitem[Ueda et al. 2005]{ued05}
%Ueda, Y., Mitsuda, K., Murakami, H., \& Matsushita, K.
%2005, ApJ, 620, 274.


\bibitem[van der Tak et al. 2003]{vdt03}
van der Tak, F.F.S., Boonman, A.M.S., Braakman, R., \& van Dishoeck, E.F.
2003, A\&A, 412, 133

\bibitem[van Dishoeck 1988]{vd88}
van Dishoeck, E. F.
1988, Rate Coefficients in Astrochemistry.  Editors, T.J. Millar, D.A. Williams; Publisher, 
Kluwer Academic Publishers, Dordrecht, Boston, 49.


\bibitem[van Zadelhoff et al. 2002]{vz02}
van Zadelhoff, G.-J., Dullemond, C. P., van der Tak, F. F. S. et al.
2002, A\&A, 395, 373.

\bibitem[Wakelam et al. 2004]{wak04}
Wakelam, V., Caselli, P., Ceccarelli, C., Herbst, E. \& Castets, A.
2003, A\&A, 422, 159


\bibitem[Wannier 1980]{wan80}
Wannier, P. G.
1980, ARA\&A, 18, 399.


\bibitem[Zhou et al. 1993]{zou93}
Zhou, S., Jaffe, D. T., Howe, J. E. et al.
1993, ApJ, 419, 190.

 
\end{thebibliography}
\end{document}